\documentclass[aps,preprint,unsortedaddress,floats]{revtex4}
\usepackage{graphicx}
\usepackage{amssymb}
\usepackage{amsmath}
\usepackage{multirow}
\usepackage{color}
\begin{document}

\title{Survivability of dust in tokamaks: dust transport in the divertor sheath}

\author{Gian Luca Delzanno and Xian-Zhu Tang}
\affiliation{Theoretical Division,
Los Alamos National Laboratory,
Los Alamos, New Mexico 87545}

\date{\today}

\begin{abstract}
  The survivability of dust being transported in the magnetized sheath near
  the divertor plate of a tokamak and its impact on the desired balance of
  erosion and redeposition for a steady-state reactor are
  investigated. Two different divertor scenarios are considered. The
  first is characterized by an energy flux perpendicular to the plate
  $q_0\simeq 1$ MW/m$^2$ typical of current short-pulse tokamaks. The
  second has $q_0\simeq 10$ MW/m$^2$ and is relevant to long-pulse
  machines like ITER or DEMO.

  It is shown that micrometer dust particles can survive rather easily
  near the plates of a divertor plasma with $q_0\simeq 1$ MW/m$^2$
  because thermal radiation provides adequate cooling for the dust
  particle. On the other hand, the survivability of micrometer dust
  particles near the divertor plates is drastically reduced when
  $q_0\simeq 10$ MW/m$^2$. Micrometer dust particles redeposit their
  material non-locally, leading to a net poloidal mass migration
  across the divertor.  Smaller particles (with radius $\sim 0.1$
  $\mu$m) cannot survive near the divertor and redeposit their
  material locally. Bigger particle (with radius $\sim 10$ $\mu$m) can
  instead survive partially and move outside the divertor strike
  points, thus causing a net loss of divertor material to dust
  accumulation inside the chamber and some non-local redeposition.
  The implications of these results for ITER are discussed.
\end{abstract}

\maketitle

\section{Introduction}

Burning plasmas ($T \sim$~10~keV or $10^8$~K) in a DT
(deuterium-tritium) fusion reactor generate intense thermal/particle
(up to tens of MW/m$^2$) and 14~MeV neutron fluxes that ultimately
must interact with a material boundary. Plasma facing components
(PFCs) must absorb and survive the enormous power and particle flux,
while maintaining structural integrity and thermo-mechanical
properties under tens of dpa neutron damage.  While the liquid metal
wall concept is currently under intensive studies, a solid wall
remains the front-runner in fusion reactor PFC design. The top
consensus material for a solid PFC is
tungsten~\cite{suzuki-etal-fst-2003,ihli-etal-fed-2006} due to its low
sputtering yield, high thermal conductivity, and high melting
temperature.  The ion bombardment flux to the PFC in the divertor
region is about $10^{23-24}$~m$^{-2}$s$^{-1}$ of $1-10^3$~eV deuterium
and tritium, and $10^{22-23}$~m$^{-2}$s$^{-1}$ of $10-10^4$~eV helium.
Without redeposition, the first wall of a tokamak would be etched away
rather quickly. For instance, gross erosion at the limiter tip of
TEXTOR was estimated at one meter per run year \cite{winter00}.
Indeed, steady-state fusion reactors rely on a near perfect local
balance between erosion and redeposition everywhere on PFCs.  The
conventional picture is that wall materials are eroded (e.g. via
sputtering) and released to the plasma as atoms and molecules, which
are ionized via charge exchange and collisional impact ionization, and
then brought back to the wall surface to redeposit by the plasma flow.
Plasma-material interaction (PMI) studies have been focused on
understanding the impurity (wall material) neutral and ion transport
and finding ways to manipulate redeposition profiles to match the
local erosion rate \cite{rudakov07}.

It is well known, however, that PMI can also create mobilizable solid
particulates or dust (from current carbon-based tokamaks, the dust
production rate relative to gross erosion is estimated around
10-15$\%$ \cite{grisolia09}), whose role in affecting the required
local balance between erosion and redeposition is particularly
important for a solid tungsten PFC. This is a far greater challenge,
as impurity neutrals can be ionized near the wall and hence promptly
sent back to the wall, but solid dust particles can traverse long
distances in the plasma chamber \cite{krash04}.  The dust particles
either stay as dust as plentifully collected on all existing
short-pulse tokamaks (causing safety issues for fusion reactors) or
redeposit non-locally, which leads to large net local erosion and thus
causes a point failure on the PFC.  For a tungsten PFC, physical and
chemical sputtering produce relatively modest erosion. Instead, helium
and hydrogen ion fluxes, coupled with a high wall temperature, cause
blistering \cite{shu07}, the formation of pits, holes, and bubbles
\cite{nishijima03,nishijima04}, and micron-in-length nano-fuzz
\cite{baldwin08}, which are sources for erosion in the form of dust
particles.

Dust particles are normally not a concern in current short-pulse tokamaks where the energy fluxes
to the walls are relatively moderate (a reference number is $q_{ref}=1$ MW/m$^2$), although occasionally a big chunk of material released by the wall
can trigger a disruption \cite{krash11}. However, long-pulse machines like ITER or DEMO are characterized by much higher energy fluxes ($q_{ref}=10$ MW/m$^2$),
stronger plasma-material interaction and are estimated to produce hundreds kg of dust. Indeed, dust safety limits are in place for ITER, setting
the maximum amount of dust that can be present at any given time in the vessel to avoid a shutdown. 
The most stringent constraint is set by dust on hot surfaces: for a design with carbon, the maximum amount is 6 kg of carbon,
 6 kg of tungsten and 6 kg of beryllium, while for a design involving only metallic materials the limit is 11 kg of beryllium and 77 kg of tungsten. Interestingly, 
Roth \textit{et al.} \cite{roth09} estimated dust production for ITER relevant conditions and concluded that in a design with carbon the dust limit on hot surfaces could be 
reached in a few tens of discharges.  These estimates, however, are based on engineering extrapolations and do not take into account the physics of dust-plasma 
interaction and dust
transport in the chamber. Therefore, a natural question is whether dust particles can survive in the plasma environment of long-pulse machines like ITER
and if indeed many kg of dust can be accumulated in such machines. Obviously this is another facet of the PMI problem: large quantities
of dust present in the machine imply a net loss of wall materials and potentially areas of net local erosion. 

In order to provide some insight on the issue of dust survivability
and its impact on the PMI problem, in this paper we study dust
transport in the divertor sheath-presheath of a tokamak.  We emphasize
an intuitively obvious fact that dust particles must stay in proximity
of the walls, where the energy fluxes are lower, to be able to survive
in long-pulse tokamaks like ITER.  This aspect differentiates our
study from other studies of dust transport in tokamaks which are more
focused on the equally important problem of core contamination
\cite{krash04, pigarov05, smirnov07, martin08, bacharis10,
  ratynskaia13} and therefore investigate dust transport outside the
sheath-presheath.

Our paper is organized as follows. In Sec. \ref{dustmodel}, we
introduce the dust transport equations used in this study. The model
consists of dust charging, heating, mass loss equations and the dust
equation of motion, in the framework of the Orbital Motion Limited (OML) theory.  
The dust transport model is coupled with a model
of the plasma sheath-presheath near the divertor plate, which we
discuss in Sec. \ref{sheathmodel}. This model is based on the
Braginskii fluid equations \cite{braginskii} and, unlike conventional
sheath models, features an equation for the conservation of energy,
such that the plasma profiles vary consistently with the plasma flow
acceleration towards the plate.  In Sec. \ref{res} we present
simulations of dust transport in the divertor sheath-presheath,
comparing and contrasting the dynamics and survivability of dust
particles in an environment characterized by $q_{ref}=1$ MW/m$^2$
relative to $q_{ref}=10$ MW/m$^2$.  Dust survivability depends on the
competition between cooling and heating. Thermal radiation is the main
cooling mechanism and is limited by the sublimation/evaporation
temperature of the dust material, $q_{rad}\sim T^4$. For tungsten, the
evaporation temperature is $T=5930$ K and $q_{rad}\sim 70$
MW/m$^2$. On the other hand, the heating energy flux associated with
electron collection scales as $q_e \sim n_e v_{th,e} T_e$. For an
electron temperature $T_e=10$ eV, $q_e \sim 21
\displaystyle{\frac{n_e}{10^{19} \, {\rm m}^{-3}}}$ MW/m$^2$. Thus,
one expects that dust particles have a better chance of survival in
current short-pulse tokamaks ($q_{ref}=1$ MW/m$^2$, $n_e\sim 10^{19}$
m$^{-3}$) where $q_e<q_{rad}$, than in long pulse-tokamaks
($q_{ref}=10$ MW/m$^2$, $n_e\sim 10^{20}$ m$^{-3}$), where
$q_e>q_{rad}$.  Our simulations confirm this qualitative picture and
place constraints on the dust size needed for survival: While in the
lower energy flux environment typical of current short-pulse tokamaks
micrometer dust particles can quite easily survive near the divertor
plates because of adequate cooling by thermal radiation, for $q_{ref}=10$
MW/m$^2$ they do not survive and redeposit their material
non-locally. Bigger particles (of radius $\sim 10$ $\mu$m), on the
other hand, can survive partially, thereby causing a net loss of divertor
material. The conclusions and the implications for long-pulse machines
like ITER or DEMO are presented in Sec. \ref{concl}.

\section{Dust transport model}\label{dustmodel}
We now discuss the model for dust transport in a plasma. As soon as the dust particle meets the plasma, it begins to charge by collection of background
plasma and electron emission \cite{shukla01}.
The subsequent dust dynamics is governed primarily by the electrostatic and drag forces. Therefore our model couples the equation of motion for the dust grain to a model for dust
grain charging. Furthermore, since the dust-plasma interaction can result in significant heat and mass sublimation/evaporation of the dust particle, the model also includes a dust heating
equation and an equation for dust mass loss. We proceed to describe each component of the model separately.

\subsection{Dust charging equation}
For dust charging, we follow the OML approach \cite{mottsmith26} 
and consider dust collection of background plasma particles and electron emission due to thermionic emission and due
to secondary emission caused by the impact of energetic plasma particles. 
The equation for the evolution of the dust charge $Q_d=e Z_d$ ($e$ is the elementary charge) is
\begin{equation}
\frac{d Q_d}{dt}=I_{i}+I_{e}+I_{se}+I_{th},
\label{qd}
\end{equation}
where $I_\alpha$ represents the various collection and emission currents defined below. The dust charge is related to the dust surface potential by 
\begin{equation}
Q_d=4 \pi \epsilon_0 r_d \phi_d\left(1+r_d/\lambda_{screen} \right),
\label{Qdscreen}
\end{equation}
with $\varepsilon_0$ the permittivity of vacuum, $r_d$ the dust radius, $\phi_d$ the dust potential, and $\lambda_{screen}$ the screening length near the grain.
In this study, we assume $\lambda_{screen}=\lambda_{lin}$ when $r_d<\lambda_{lin}$, where
the linearized Debye length is given by
$1/\lambda_{lin}^2=1/\lambda_{De}^2+1/\lambda_{Di}^2$ [with the electron and ion Debye lengths defined by 
$\lambda_{De,i}=\sqrt{\varepsilon_0 T_{e,i}/\left(e^2 n_{e,i}\right)}$, $T_{e,\,(i)}$ and $n_{e,\,(i)}$ are the local electron (ion)
temperature and density],
while for $r_d>\lambda_{lin}$ we use $\lambda_{screen}=\lambda_{De}$. This is a rough attempt to take into account the results of
Ref.  \cite{daugherty92}, showing that the screening length increases with the dust radius.
In Eq. (\ref{qd}),
the ion collection current is
\begin{eqnarray}
&& I_{i}=e \pi r_d^2 n_{i} \sqrt{\frac{8 T_{i}}{\pi m_{i}}}\left[\frac{\sqrt{\pi}}{4 u} 
\left(1+2 u^2-2 \frac{e \phi_d}{T_i} \right) {\rm erf} \left( u \right) +\frac{1}{2} \exp \left(-u^2 \right)\right],\,\,\,\phi_d<0 \nonumber \\
&& I_{i}=e \pi r_d^2 n_{i} \sqrt{\frac{8 T_{i}}{\pi m_{i}}} \Bigg\{ \frac{\sqrt{\pi}}{8 u}
\left(1+2 u^2-2 u_m^2 \right) \left[{\rm erf} \left( u-u_m \right)+ {\rm erf} \left( u+u_m \right)\right]+
\nonumber \\
&& \,\,\,\,\,\,\frac{1}{4}\left(1+\frac{u_{m}}{u} \right) \exp \left[-\left( u-u_{m}\right)^2 \right]+
\frac{1}{4}\left(1-\frac{u_{m}}{u} \right) \exp \left[-\left( u+u_{m}\right)^2 \right]\Bigg\},\,\,\,\phi_d>0, \nonumber \\
\label{ipa}
\end{eqnarray}
where  $m_i$ is the ion mass,
\begin{equation}
u=\displaystyle{\frac{|{\bf V}_{i}-{\bf V}_d|}{v_{thi}}}
\end{equation}
and
\begin{equation}
u_{m}=\sqrt{\frac{e \phi_d}{T_{i}}},\,\,\,\phi_d>0.
\end{equation}
The ion thermal velocity is $v_{thi}=\sqrt{\frac{2 T_{i}}{m_{i}}}$, while ${\bf V}_i$ and $\bf{V}_d$ are the ion and dust flow, respectively.
The electron collection current is
\begin{eqnarray}
&& I_e=-e \pi r_d^2 n_e \sqrt{\frac{8 T_e}{\pi m_e}}\exp \left( \frac{e \phi_d}{T_e}\right),\,\,\,\phi_d<0 \nonumber \\
&& I_e=-e \pi r_d^2 n_e \sqrt{\frac{8 T_e}{\pi m_e}} \left(1+ \frac{e \phi_d}{T_e}\right),\,\,\,\phi_d>0 \label{ie},
\end{eqnarray}
with $m_e$ the electron mass.
From Eq. (\ref{ipa}), one can notice that the ion collection current accounts for the fact that the ions have a finite mean velocity,
since such velocity can be comparable to the ion thermal speed in the sheath-presheath. 
%On the other hand, the electron thermal velocity
%is normally higher than the sonic sheath flow, therefore electron current collection is described by OML without accounting for a finite drift.
The secondary emission current due to electron impact (ion impact is usually neglected since it becomes important for ion energies above $\sim 1$ keV) is
given by \cite{shukla01}
\begin{eqnarray}
&& I_{se}=e\frac{8 \pi^2 r_d^2}{m_e^2}\int_0^{\infty}E \delta_{se}\left(E\right)f_{se}\left(E-e\phi_d\right)dE,\,\,\,\phi_d<0 \nonumber \\
&& I_{se}=e\frac{8 \pi^2 r_d^2}{m_e^2}\exp \left(-\frac{e \phi_d}{T_{se}}\right)\left(1+\frac{e \phi_d}{T_{se}} \right)
\int_{e \phi_d}^{\infty}E \delta_{se}\left(E\right)f_{se}\left(E-e\phi_d\right)dE,\,\,\,\phi_d>0, \nonumber \\ \label{ise}
\end{eqnarray}
where $E$ is the kinetic energy of the incident electron, $T_{se}$ is the temperature of the secondary emitted electrons, the secondary emission yield is given by the Sternglass formula \cite{Bruining}
\begin{equation}
\delta_{se}(E)=\frac{7.4 \delta_m E}{E_m}\exp\left(-2 \sqrt{\frac{E}{E_m}} \right)
\label{yield}
\end{equation}
with $\delta_m$ and $E_m$ parameters that depend on the dust material ($\delta_m=2.4$ and $E_m=400$ eV for silicates), and
\begin{equation}
f_{se}\left(E-e\phi_d \right)=n_e \left(\frac{m_e}{2 \pi T_e} \right)^{3/2} \exp \left(-\frac{E-e\phi_d}{T_e} \right)
\label{fse}
\end{equation}
is the distribution function of the incident electrons.
The thermionic current is given by the Richardson-Dushman formula \cite{ashcroft, sodha71}
\begin{eqnarray}
&& I_{th}=e\frac{16 \pi^2 r_d^2 m_e T_d^2}{h^3}\exp \left(-\frac{W}{T_d} \right),\,\,\,\phi_d<0 \nonumber \\
&& I_{th}=e\frac{16 \pi^2 r_d^2 m_e T_d^2}{h^3}\left(1+\frac{e \phi_d}{T_d}\right)\exp \left(-\frac{W+e\phi_d}{T_d} \right),\,\,\,\phi_d>0, \label{ith}
\end{eqnarray}
where $T_d$ is the dust surface temperature, $h$ is Planck's constant, and $W$ the thermionic work function of the dust material.
We note that, for a positively charged grain, the OML formulas do not take into account the fact that a potential well can form near
the grain, as shown in Refs. \cite{delzanno04,delzanno05}.

\subsection{Dust equation of motion}
The dust charging model is coupled to the equation of motion of the dust grain
\begin{eqnarray}
&& \frac{d {\bf x}_d}{dt}={\bf V}_d \nonumber \\
&& m_d \frac{d {\bf V}_d}{dt}=Q_d \left({\bf E}+{\bf V}_d \times {\bf B}\right)+m_d {\bf g}+{\bf F}_{id},
\label{eom}
\end{eqnarray}
where $m_d=4 \pi r_d^3 \rho_d/3$ is the dust mass ($\rho_d$ is the dust density),
${\bf E}$ and ${\bf B}$ are the local electric and magnetic field, and {\bf g} is gravity ($|{\bf g}|=9.8 $ m/s$^2$).
We note that in Eq. (\ref{eom}) we do not include the rocket force which can be present when the dust particle is losing mass,
thus implicitly assuming that the dust material is lost with zero velocity relative to the dust particle.

The ion drag force can be calculated in the framework of the OML theory. 
It is normally divided between the contribution to the drag by the ions that are directly collected by the dust grain (labeled as ${\bf F}_{id,\,coll}$),
and the scattering part due to the Coulomb interaction between the dust grains and the ions (not collected by the grain) orbiting in the dust grain sheath (labeled as ${\bf F}_{id,\,orb}$).
For a negatively charged grain,
the drag collection is given by
\begin{equation}
{\bf F}_{id,\,coll}=\pi r_d^2 m_i n_i v_{thi} \left[\frac{1}{\sqrt{\pi}}\left(1+2 w_{-}\right)\exp\left(-u^2\right)+
u\left(1+2 w_{-} - \frac{1-2 w_{+}}{2 u^2}\right){\rm erf}\left(u\right) \right] \frac{{\bf V}_i-{\bf V}_d}{2 u^2}
,\,\,\,\phi_d<0,
\label{Fidcollneg}
\end{equation}
while for a positively charged grain it is given by
\begin{eqnarray}
&& {\bf F}_{id,\,coll}=\pi r_d^2 m_i n_i v_{thi} \Bigg\{ \frac{1}{\sqrt{\pi}}
\left[\left(1+2 u^2+\frac{1-2 u^2}{u}u_m\right)\exp\left[-\left(u+u_m\right)^2\right]\right.+ \nonumber \\
&& \,\,\,\,\,\,\left.\left(1+2 u^2-\frac{1-2 u^2}{u}u_m\right)\exp\left[-\left(u-u_m\right)^2\right]\right]+ \nonumber \\
&& \,\,\,\,\,\,u\left(1+2 w_{-} - \frac{1-2 w_{+}}{2 u^2}\right)\left[{\rm erf}\left(u+u_m\right)+{\rm erf}\left(u-u_m\right) \right] \Bigg\}\frac{{\bf V}_i-{\bf V}_d}{4 u^2},\,\,\,\phi_d>0.
\label{Fidcollpos}
\end{eqnarray}
In Eqs. (\ref{Fidcollneg}) and (\ref{Fidcollpos}) we have defined $w_{\pm}=u^2\pm e\phi_d/T_i$.
For the orbital part of the ion drag force, we follow the model described in Refs. \cite{pigarov05,smirnov07}. That is,
for both negatively and positively charged grains, we have
\begin{equation}
{\bf F}_{id,\,orb}=2 \pi r_d^2 m_i n_i v_{thi} \left(\frac{e\phi_d}{T_i}\right)^2\frac{\mathcal{G}\left(u\right)}{u}\log \Lambda \left({\bf V}_i-{\bf V}_d\right),
\label{Fidorb}
\end{equation}
where the Chandrasekhar function is $\mathcal{G}(u)=\left[{\rm erf}\left(u\right)-2 u \exp\left(-u^2\right)/\sqrt{\pi}\right]/\left(2u^2\right)$. For a negatively charged grain, the Coulomb logarithm $\log \Lambda$
is calculated as
\begin{equation}
\log \Lambda= \frac{1}{2} \log \frac{b_{90}^2+\eta_{fit}^2 \lambda_s^2}{b_{90}^2+r_d^2},\,\,\,\,\phi_d<0,
\label{coullogneg}
\end{equation}
with the impact parameter $b_{90}=-\displaystyle{\frac{r_d}{3+2 u^2}\frac{e \phi_d}{T_i}}$, the screening length
$\lambda_s=\lambda_{De}/\sqrt{1+\displaystyle{\frac{3 T_e}{T_i \left(3+2u^2 \right)}}}$
and $\eta_{fit}=1+\frac{r_d}{\lambda_s}\left(1+\sqrt{\frac{T_e}{6 T_i}} \right)$
used in Ref. \cite{pigarov05,smirnov07} to fit PIC simulation results obtained by Hutchinson \cite{hutchinson03}.
For a positively charged grain, the Coulomb logarithm is given by \cite{khrapak04}
\begin{equation}
\log \Lambda=\int_0^{+\infty} \exp(-x) \log \left[1+4  \left( \frac{\lambda_{lin}}{r_d} 
\frac{T_i}{e \phi_d} x\right)^2\right]dx-2\int_{e \phi_d/T_i}^{+\infty} \exp(-x) 
\log\left(2\frac{T_i}{e \phi_d}x-1 \right)dx,\,\,\,\phi_d>0.
\label{coullogpos}
\end{equation}
%which reduces to
%\begin{equation}
%\log \Lambda \simeq 2 \left[\log \left(2 \frac{\lambda_{lin}}{r_d} \frac{T_i}{e \phi_d}\right) 
%-0.577+\exp\left(-\frac{e \phi_d}{2 T_i}\right){\rm E}_{{\rm i}}\left(-\frac{e \phi_d}{2 T_i} \right)\right]
%\label{approxcoullog}
%\end{equation}
%in the limit $\lambda_{lin} T_i /\left(e \phi_d r_d \right)\gg 1$. 
In the sheath-presheath of a tokamak, an estimate of the forces acting on the micrometer
dust grains indicates that the electrostatic force and the ion drag are dominant, while the magnetic part of the Lorentz force and gravity are negligible \cite{krash04,tang10}.
Gravity becomes important for bigger dust grains of radius $r_d\sim 100$ $\mu$m.

\subsection{Dust heating equation}
As the dust grain interacts with the plasma, it can heat up substantially. This gives rise to a considerable thermionic emission current, which changes the dust
floating potential and affects the current collection and the drag forces on the grain, and, if conditions for evaporation/sublimation are met, 
can lead to mass loss. We consider the following dust heating equation 
\begin{equation}
C_d \frac{d (m_d T_d)}{d t}= q_{e} +q_{i}-q_{se}-q_{th} - q_{rad}+q_{rec},
\label{heat}
\end{equation}
where $C_d$ is the specific heat capacity of the grain, assumed to be constant. 
In Eq. (\ref{heat}), the energy fluxes associated with the background plasma collection and electron emission are calculated from OML theory.
The energy flux resulting from ion collection is
\begin{eqnarray}
q_i&=&\sqrt{\pi}r_d^2 n_i T_i \sqrt{\frac{2 T_i}{m_i u^2}}\Bigg\{u \left(\frac{5}{2}+u^2-\frac{e \phi_d}{T_i}\right)\exp\left(-u^2\right)+ \nonumber \\
&& \sqrt{\pi}\left[\frac{3}{4}+3u^2+u^4-\frac{e \phi_d}{2 T_i} \left(1 +2u^2\right) \right]{\rm erf} \left( u \right)\Bigg\},\,\,\,\phi_d \le 0, \nonumber \\
q_i&=&\sqrt{\pi}r_d^2 n_i T_i \sqrt{\frac{2 T_i}{m_i u^2}}\Bigg\{A_+ + A_-+\frac{\sqrt{\pi}}{2}\left[\frac{3}{4}+3u^2+u^4-\frac{u_m^2}{2}\left(1+2u^2 \right) \right] \nonumber \\
&& \left[ {\rm erf}\left(u+u_m \right)+{\rm erf}\left(u-u_m \right)\right]\Bigg\}, \,\,\,\phi_d > 0,
\label{gami}
\end{eqnarray}
where
\begin{eqnarray}
A_{\pm}=u\left(\frac{5}{4}+\frac{u^2}{2}\mp \frac{3+2 u^2}{4 u}u_m \right) \exp\left[-\left(u \pm u_m\right)^2\right].
\end{eqnarray}
For electron collection and thermionic/secondary emission we have
\begin{eqnarray}
&& q_{e,\,th,\,se}=2 T_{e,\,d,\,se} \frac{|I_{e,\,th,\,se}|}{e}, \,\,\,\phi_d \le 0, \nonumber \\
&& q_{e,\,th,\,se}=T_{e,\,d,\,se} \left(\frac{\displaystyle{2+\frac{e \phi_d}{T_{e,\,d,\,se}}}}{\displaystyle{1+\frac{e \phi_d}{T_{e,\,d,\,se}}}} + 
\frac{e \phi_d}{T_{e,\,d,\,se}}\right)\frac{|I_{e,\,th,\,se}|}{e}, \,\,\,\phi_d > 0.
\label{game}
\end{eqnarray}
The thermal radiation flux is given by the Stefan-Boltzmann law
\begin{equation}
q_{rad}=4 \pi r_d^2 \varepsilon \sigma \left(T_d^4-T_{w}^4\right),
\label{gamrad}
\end{equation}
where $\varepsilon$ is the emissivity of the dust material (which depends on dust size and temperature), 
$\sigma$ is the Stefan-Boltzmann constant and $T_{w}$ is the wall temperature.
Finally, since the dust grain acts as a site where electrons and ions can recombine and a neutral atom or molecule can be released,
we account for the energy flux associated with such process (see the discussion in Refs. \cite{martin08, bacharis10})
\begin{equation}
q_{rec}=\left(13.6e+1.1e-T_{d}\right)\frac{I_i}{e}.
\label{qrec}
\end{equation}
Equation (\ref{qrec}) assumes that hydrogen/deuterium molecules are released.

\subsection{Dust mass loss equation}
When conditions for evaporation or sublimation are met, the dust grain loses mass according to the following equation
\begin{equation}
\frac{d m_d}{dt}=-\frac{q_{net}}{H},
\label{massloss}
\end{equation}
where $q_{net}$ is the net energy flux on the grain (sum of all the energy fluxes described above), 
and $H$ is the latent heat for the relevant phase change. We note that Eq. (\ref{massloss})
is applied only for $q_{net}>0$, since $q_{net}<0$ would correspond to the grain collecting mass. For a spherical dust grain
and assuming isotropic mass loss, Eq. (\ref{massloss})
becomes an equation for the evolution of the dust radius:
\begin{equation}
4 \pi r_d^2 \rho_d \frac{d r_d}{dt}=-\frac{q_{net}}{H}.
\label{radiusloss}
\end{equation}

Equations (\ref{qd}), (\ref{eom}), (\ref{heat}) and (\ref{radiusloss}) constitute the model that will be used in the Sec. \ref{res} to study dust
transport in the magnetized sheath-presheath of the divertor. Overall this model is quite similar to other models of dust transport in a plasma,
including the ones that are currently used for tokamak applications \cite{ticos06,pigarov05,smirnov07,martin08,bacharis10,ratynskaia13}.

\section{One-dimensional model of the magnetized sheath-presheath} \label{sheathmodel}

The model of dust transport described in the previous section is coupled to a model for the local conditions of the background plasma and electromagnetic field.
Since we are interested in dust transport near the walls, where chances of survival are maximized, we develop a model for the magnetized sheath-presheath
near the divertor plates.
We consider a system where a plasma interacts with a planar divertor plate lying horizontally. We use Cartesian geometry $(x,y,z)$, where $x$ is the direction
perpendicular to the plate. A stationary magnetic field 
${\bf B}=(B_x,B_y,B_z)=B_0 (\sin \theta \cos\psi,\cos\theta\cos\psi,\sin\psi)$ is present, 
with $\theta$ and $\psi$ the angles relative to the plate and $B_0$ a reference magnetic field. 
The system is one-dimensional and all the quantities depend only on $x$. Since we consider $\theta,\psi \ll 90^\circ$, the $y$ direction
corresponds to the toroidal direction of a tokamak, while $z$ is the poloidal direction.

The goal of this section is to build a sheath-presheath model within the framework of a fluid model for the plasma that includes an energy equation.
In order to do this, we start from the Braginskii equations for a collisional plasma
consisting of electrons and singly charged ions \cite{braginskii}.
The electron and ion collisional times are defined as
\begin{equation}
\tau_e=\frac{6 \sqrt{2} \pi^{3/2} \varepsilon_0^2 \sqrt{m_e} T_e^{3/2}}{\log \Lambda e^4 n_i},\,\,\, 
\tau_i=\frac{12 \pi^{3/2} \varepsilon_0^2 \sqrt{m_i} T_i^{3/2}}{\log \Lambda e^4 n_i}, \label{tau}
\end{equation}
where $\log \Lambda \simeq 6.6 -0.5 \log \frac{n_e}{10^{20} m^{-3}}+1.5 \log \frac{T_e}{1\,{\rm eV}}$.
The Coulomb logarithm is a very slowly varying function of density and temperature and it will be assumed
constant in the sheath-presheath model.

We consider the steady-state Braginskii equations in the magnetized
limit corresponding to $\omega_{ce}\tau_e,\,\omega_{ci}\tau_i\gg 1$,
with the electron (ion) cyclotron frequency given by
$\omega_{ce(i)}=e B_0/m_{e(i)}$.  Neglecting terms due to the
stress-tensor, these equations, in normalized units to be described
below, are:
\begin{eqnarray}
&& \nabla \cdot \left(n_e {\bf V}_e \right)=\nu_i n_e \label{ne} \\
&& \frac{m_e}{m_i} n_e \left({\bf V}_e\cdot \nabla \right){\bf V}_e=-\alpha \nabla p_e -\alpha n_e {\bf E} -n_e{\bf V}_e \times {\bf B}+{\bf R}-\frac{m_e}{m_i}\nu_i n_e {\bf V}_e \label{ve} \\
&& \frac{3}{2}n_e \left({\bf V}_e \cdot \nabla \right)T_e+p_e \nabla \cdot {\bf V}_e+\nabla \cdot {\bf q}_e=Q_e+
\frac{m_e}{m_i}\frac{\nu_i}{2 \alpha} n_e {\bf V}_e^2+\frac{3}{2} \nu_i n_e \left(T_n^e-T_e\right) \label{pe} \\
&& \nabla \cdot \left(n_i {\bf V}_i \right)=\nu_i n_e \label{ni} \\
&& n_i \left({\bf V}_i\cdot \nabla \right){\bf V}_i=-\alpha \nabla p_i +\alpha n_i{\bf E}+n_i {\bf V}_i \times {\bf B}-{\bf R}-\nu_i n_e {\bf V}_i \label{vi} \\
&& \frac{3}{2} n_i \left({\bf V}_i \cdot \nabla \right)T_i+p_i \nabla \cdot {\bf V}_i+\nabla \cdot {\bf q}_i=Q_i+
\frac{\nu_i}{2 \alpha}n_e {\bf V}_i^2+\frac{3}{2} \nu_i n_e \left(T_n^i-T_i\right),
\label{pi}
\end{eqnarray}
for the electron density $n_e$, velocity field ${\bf V}_e$ and pressure $p_{e}=n_{e} T_{e}$ and similarly for the ions.
Moreover, the electric field ${\bf E}=-\nabla \phi$ is obtained by Poisson's equation
\begin{equation}
\nabla^2 \phi= \left(\frac{\rho_i}{\lambda_{De}}\right)^2\left(n_e-n_i \right).
\label{pois}
\end{equation}
Equations (\ref{ne})-(\ref{pois}) have been written according to the following normalization:
${\bf x} \longrightarrow {\bf x}/\rho_i$, ${\bf V}_{e,i} \longrightarrow {\bf V}_{e,i}/C_s$, 
$T_{e,i} \longrightarrow T_{e,i}/T_{e0}$, $n_{e,i} \longrightarrow n_{e,i}/n_0$, 
$p_{e,i}\longrightarrow p_{e,i}/\left(n_0 T_{e0} \right)$, $\phi \longrightarrow e\phi/T_{e0}$,
$ {\bf B} \longrightarrow {\bf B}/B_0$, $\nu_i \longrightarrow \nu_i/\omega_{ci}$, 
where $n_0$, and $T_{e0}$ ($T_{i0}$) are some reference density and electron (ion) temperature.
We have defined the ion sound speed as $C_s=\sqrt{\left(T_{e0}+T_{i0}\right)/m_i}$, the ion gyroradius as
$\rho_i=C_s/\omega_{ci}$, 
and $\alpha=T_{e0}/\left(T_{e0}+T_{i0}\right)$. We note that in the continuity equations (\ref{ne}) and (\ref{ni}) we have included a source term due to ionization.
We assume a constant ionization frequency, $\nu_i$, implying that we are implicitly assuming a stationary and uniform background of neutrals.
We have not attempted to model accurately the rather complex energetics of the ionization process. For this, we have simply introduced
two fictitious temperatures, $T_n^e$ and $T_n^i$, which can be varied parametrically. We also note that the form of the terms proportional to $\nu_i$ in
the momentum and energy equations is due to the fact that these equations are written in non-conservative form.

In Eqs. (\ref{ne})-(\ref{pi}), the collisional term in the momentum equations, ${\bf R}$, is composed of the friction force due to the relative motion
between electrons and ions and of the thermal force:
\begin{equation}
{\bf R}=-\frac{1}{\omega_{ce}\tau_e}n_e \left[0.5\left({\bf V}_{e\parallel}-{\bf V}_{i\parallel}\right)+ 
{\bf V}_{e\perp}-{\bf V}_{i\perp}\right]-0.7\alpha n_e \nabla_\parallel T_e
-\frac{3}{2}\frac{\alpha}{\omega_{ce}\tau_e}n_e {\bf b}\times \nabla_\perp T_e,
\label{R}
\end{equation}
where $\parallel$ and $\perp$ refer to the direction parallel and perpendicular to the magnetic field
and ${\bf b}={\bf B}/|{\bf B}|$ is the unit vector along ${\bf B}$.
The electron and ion heat fluxes are given, respectively, by
\begin{equation}
{\bf q}_e=-\chi^e_\parallel \nabla_\parallel T_e-\chi^e_\perp \nabla_\perp T_e-\chi^e_\times {\bf b}\times \nabla_\perp T_e
+0.7 n_e T_e \left({\bf V}_{e\parallel}-{\bf V}_{i\parallel}\right)+\frac{3}{2}\frac{n_e T_e}{\omega_{ce}\tau_e}
{\bf b} \times \left({\bf V}_e-{\bf V}_i \right)
\label{qe}
\end{equation}
and
\begin{equation}
{\bf q}_i=-\chi^i_\parallel \nabla_\parallel T_i-\chi^i_\perp \nabla_\perp T_i+\chi^i_\times {\bf b}\times \nabla_\perp T_i
\label{qi}
\end{equation}
where the thermal conductivities are
\begin{eqnarray}
&& \chi^e_\parallel=3.2 \alpha \omega_{ce} \tau_e n_e T_e,\,\,\,\chi^e_\perp=4.7 \frac{\alpha}{\omega_{ce} \tau_e} n_e T_e,\,\,\, \chi^e_\times=2.5 \alpha n_e T_e \label{chie} \\
&& \chi^i_\parallel=3.9 \alpha \omega_{ci}\tau_i n_i T_i,\,\,\,\chi^i_\perp=2 \frac{\alpha}{\omega_{ci} \tau_i} n_i T_i,\,\,\,\chi^i_\times=2.5 \alpha n_i T_i. \label{chii}
\end{eqnarray}
The heat generated by the electrons in collisions with the ions is
\begin{equation}
Q_e=-{\bf R}\cdot \left({\bf V}_e-{\bf V}_i\right)-Q_i
\label{Qe}
\end{equation}
while the heat acquired by the ions is
\begin{equation}
Q_i=\frac{3}{\omega_{ce}\tau_e}n_e \left(T_e-T_i\right).
\label{Qi}
\end{equation}

The important point is that model (\ref{ne})-(\ref{pois}) features an equation for the conservation of the energy flux.
By introducing the convective energy fluxes as
\begin{equation}
{\bf q}_{e,\,i}^{\rm conv}={\bf V}_{e,\,i}\left(\frac{5}{2}p_{e,\,i}+\frac{m_{e,\,i}}{m_i}\frac{1}{2\alpha}n_{e,\,i}{\bf V}_{e,\,i}^2\right), \label{qeconv}
\end{equation}
and starting from the conservative form of the energy equations (\ref{pe}) and (\ref{pi}), it is easy to show that
\begin{equation}
\nabla \cdot \left( {\bf q}_e+{\bf q}_e^{\rm conv} + {\bf q}_i+{\bf q}_i^{\rm conv}\right)=\frac{3}{2} \nu_i n_e \left(T_n^i+T_n^e \right).
\label{energy}
\end{equation}
%In the derivation of Eq. (\ref{energy}), we have used the conservation of the current, $\nabla \cdot \left(n_i {\bf V}_i-n_e {\bf V}_e \right)=0$,
%easily obtained from the continuity equations (\ref{ne}) and (\ref{ni}).

We further manipulate Eqs. (\ref{ne})-(\ref{pois}) by exploiting 
\begin{equation}
\omega_{ce}\tau_e=\overline{\omega_{ce}\tau_e}\frac{T_e^{3/2}}{n_e},\,\,\,\overline{\omega_{ce}\tau_e}
=\omega_{ce}\frac{6 \sqrt{2} \pi^{3/2} \varepsilon_0^2 \sqrt{m_e} T_{e0}^{3/2}}{\log \Lambda_0 e^4 n_0}
\label{taubar}
\end{equation}
(where the Coulomb logarithm is evaluated with the reference normalization parameters) and similarly for the ions
and keeping only the zeroth and first order terms in $\overline{\omega_{ce,i}\tau_{e,i}}$. The resulting equations are
\begin{eqnarray}
&& \nabla \cdot \left(n_e {\bf V}_e \right)=\nu_i n_e \label{ne2} \\
&& \frac{m_e}{m_i} n_e \left({\bf V}_e\cdot \nabla \right){\bf V}_e=-\alpha \nabla p_e -\alpha n_e {\bf E} -n_e{\bf V}_e \times {\bf B}-
	0.7\alpha n_e \nabla_{\parallel}T_e-\frac{m_e}{m_i}\nu_i n_e {\bf V}_e \label{ve2} \\
&& \frac{3}{2}n_e \left({\bf V}_e \cdot \nabla \right)T_e+p_e \nabla \cdot {\bf V}_e-3.2\alpha \nabla \cdot \left(\overline{\omega_{ce}\tau_e} \frac{n_e}{n_i}
T_e^{5/2}\nabla_{\parallel}T_e\right)-2.5\alpha\nabla\cdot \left( n_e T_e {\bf b} \times \nabla T_e\right) \nonumber \\
&&\,\,\,\,\,\,+\nabla \cdot \left[ 0.7 n_e T_e \left( {\bf V}_{e\parallel}-{\bf V}_{i\parallel}\right)\right]=0.7 n_e \left({\bf V}_e-{\bf V}_{i} \right)\cdot \nabla_\parallel T_e+
\frac{m_e}{m_i}\frac{\nu_i}{2 \alpha} n_e {\bf V}_e^2+\frac{3}{2} \nu_i n_e \left(T_n^e-T_e\right) \nonumber \\ \label{pe2} \\
&& \nabla \cdot \left(n_i {\bf V}_i \right)=\nu_i n_e \label{ni2} \\
&& n_i \left({\bf V}_i\cdot \nabla \right){\bf V}_i=-\alpha \nabla p_i +\alpha n_i{\bf E}+n_i {\bf V}_i \times {\bf B}+0.7\alpha n_e \nabla_\parallel T_e-\nu_i n_e {\bf V}_i \label{vi2} \\
&& \frac{3}{2} n_i \left({\bf V}_i \cdot \nabla \right)T_i+p_i \nabla \cdot {\bf V}_i-3.9\alpha \nabla \cdot \left(\overline{\omega_{ci}\tau_i} T_i^{5/2} \nabla_\parallel T_i\right)
+2.5\alpha \nabla \cdot \left( n_i T_i {\bf b} \times \nabla T_i\right)=\nonumber \\
&&\,\,\,\,\,\,\frac{\nu_i}{2 \alpha}n_e {\bf V}_i^2+\frac{3}{2} \nu_i n_e \left(T_n^i-T_i\right), \label{pi2} \\
&& \nabla^2 \phi=\left(\frac{\rho_i}{\lambda_D}\right)^2\left(n_e-n_i\right). \label{pois2}
\end{eqnarray}
Equations (\ref{ne2})-(\ref{pois2}) are the sheath-presheath model focus of this paper. 
%It is interesting to note that 
%the collisional term in the momentum equations (\ref{ve2}) and (\ref{vi2}) is dominated by the thermal
%contribution. It is quite obvious that in the limit $\overline{\omega_{ce,i}\tau_{e,i}} \longrightarrow \infty$ (strongly magnetized limit), 
%the energy equations (\ref{pe2}) and (\ref{pi2}) essentially
%mantain the temperature profiles fairly constant. Furthermore, in the limit of zero mass ratio, the electron
%momentum equation (\ref{ve2}) essentially recovers the Boltzmann expression for the electron density.
%Thus, it easy to see that Eqs. (\ref{ne2})-(\ref{pois2}) converge to the model of Ref. \cite{riemann94} in the
%strongly magnetized limit.
Very often in the literature a distinction is made between the sheath and the presheath. The presheath is essentially a quasi-neutral
region where a weak electric field is setup to accelerate the ions to the sound speed at the entrance of sheath. This is known as Bohm's condition for
the existence of a stationary (non-oscillatory) sheath. When a magnetic field is present (and the ions are fully magnetized), 
the presheath can be further divided into the regular presheath and the magnetic presheath
(or Chodura layer). The transition between presheath and Chodura layer is reached when the parallel flow becomes sonic \cite{chodura82}.
The sheath, on the other hand, is a region of strong non-neutrality in the plasma, characterized
by a strong electric field. Thus, the physical picture of the magnetized sheath is the following: in the presheath, the ion flow is accelerated to the sound speed parallel to
the magnetic field; in the Chodura layer, the ion flow is deflected from parallel to the magnetic field to perpendicular to the wall (and equal to the sound speed); in the sheath,
a strong electric field is created to equalize the plasma currents to the wall. 
%We note that in principle there is no need to distinguish between sheath and presheath, and 
%Eqs. (\ref{ne2})-(\ref{pois2}) can be used to model the whole system.

We solve Eqs. (\ref{ne2})-(\ref{pois2}) numerically as a set of
ordinary differential equations (ODEs) by specifying the boundary conditions upstream (here upstream is relative to the position of the plate).
We use the following upstream boundary conditions: zero particle flux (implying zero convective energy flux) and
a finite conductive heat flux (namely a finite temperature gradient) for both electrons and ions, which is used to set the desired total
energy flux in the direction perpendicular to the plate (labeled $q_0$).
We note that in principle one can integrate the system of ODEs indefinitely starting from upstream.
Some conditions is therefore needed to stop the integration and determine the plate position and the system size. In our study this condition is given
by the electron velocity perpendicular to the plate equal to the local electron thermal velocity, since this corresponds to a singularity of the
sheath-presheath model, as done in Ref. \cite{duarte11}. 

\begin{figure}
\centering
\includegraphics[scale=0.8]{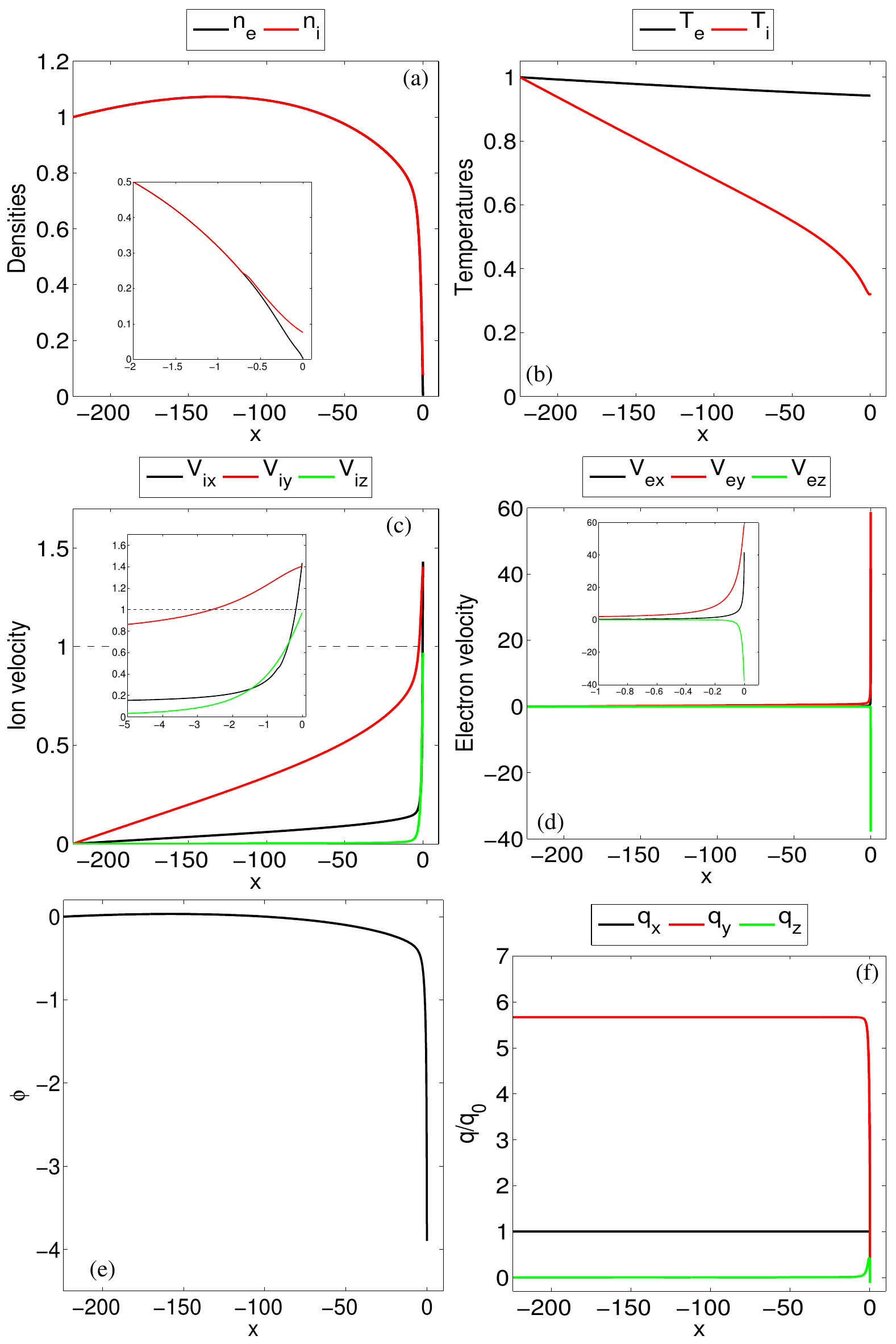}
\caption{(Color online) Braginskii sheath-presheath model with upstream parameters $n_0=2\cdot 10^{19}$ part/m$^3$, $T_{e0}=T_{i0}=10$ eV,
and $q_0=0.91$ MW/m$^2$. Plotted quantities are: (a) plasma densities, (b) plasma temperatures, (c) ion velocity components,
(d) electron velocity components, (e) electrostatic potential, (f) energy flux components. In Fig. (c) the dashed line indicates a sonic flow.}
\label{fig1}
\end{figure}

\begin{figure}
\centering
\includegraphics[scale=0.8]{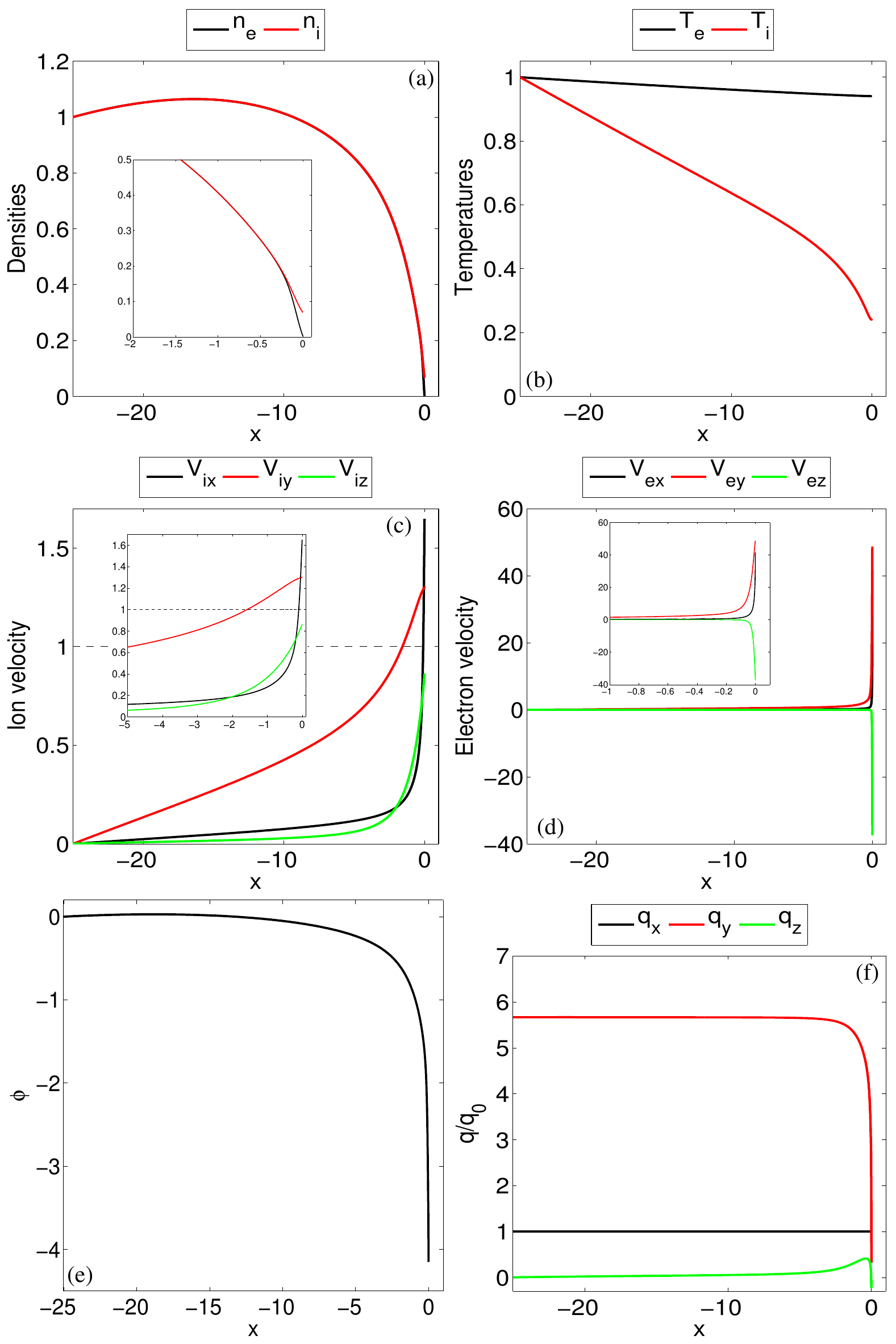}
\caption{(Color online) Braginskii sheath-presheath model with upstream parameters $n_0=2\cdot 10^{20}$ part/m$^3$, $T_{e0}=T_{i0}=10$ eV,
and $q_0=9.6$ MW/m$^2$. Plotted quantities are: (a) plasma densities, (b) plasma temperatures, (c) ion velocity components,
(d) electron velocity components, (e) electrostatic potential, (f) energy flux components. In Fig. (c) the dashed line indicates a sonic flow.}
\label{fig1b}
\end{figure}

For the results presented in this paper, we consider a deuterium plasma and a magnetic field of magnitude $B_0=6$ T, with angles relative to the plate
$\theta=\psi=10^\circ$. We also set $T_n^i=T_n^e=0$.
The conductive heat flux upstream is $q_{0e}=0.8 q_0$, while for the ions $q_{0i}=0.2 q_0$, and we adjust $q_0$
to have an ion temperature profile which is monotonically decreasing towards the plate. The ionization rate is
$\nu_i=7 \cdot 10^{-15} n_0/\omega_{ci}$. In order to study dust transport in regimes that are relevant to current short-pulse and next generation
long-pulse tokamaks, we consider two classes of equilibria.
The first is characterized by an energy flux perpendicular to the divertor plate $q_0=0.91$ MW/m$^2$ in a plasma
with upstream plasma density $n_0=2 \cdot 10^{19}$ part/m$^3$ and temperature $T_{e0}=T_{i0}=10$ eV, and is plotted in Fig. \ref{fig1}. 
The second equilibrium has
an energy flux perpendicular to the wall $q_{0}=9.6$ MW/m$^2$ in a plasma with $n_0=2 \cdot 10^{20}$ part/m$^3$ and 
$T_{e0}=T_{i0}=10$ eV, and is plotted in Fig. \ref{fig1b}.
Qualitatively the two profiles are very similar, the most obvious difference being the system size: $L/\rho_i\simeq 224$ in Fig. \ref{fig1}
and  $L/\rho_i\simeq 25$ in Fig. \ref{fig1b}. This difference is due to the ionization rate, which sets the presheath width, and is $10$ times
higher in Fig. \ref{fig1b}. 
We note that, for these parameters, the system size is comparable to the ion collision mean free path ($\lambda_i=v_{thi}/\tau_i$, $L/\lambda_i\simeq1.02$
and $L/\lambda_i\simeq0.74$ for Figs. \ref{fig1} and \ref{fig1b}, respectively) and smaller than the ionization mean free path
($\lambda_{ioniz}=v_{thi}/\nu_i$, $L/\lambda_{ioniz}\simeq0.18$
and $L/\lambda_{ioniz}\simeq0.23$).
Figures \ref{fig1} and \ref{fig1b} show the typical behavior of the magnetized sheath outlined above: 
the density decreases significantly towards the plate (a) while the plasma is accelerated,
and the ion velocity parallel to the magnetic field (which in practice coincides with $V_{iy}$) becomes sonic before the velocity
perpendicular to the plate does, marking the formation of the Chodura layer and of the sheath. In Fig. \ref{fig1} the width of the Chodura layer
is $\Delta_{cl}/\rho_i\sim 3$ while the sheath width is $\Delta_{sh}/\lambda_{De}\sim 4$. For the parameters of Fig. \ref{fig1b} we have
$\Delta_{cl}/\rho_i\sim 6$ and $\Delta_{sh}/\lambda_{De}\sim 2$. We note that the electron velocity (d) becomes much higher than sonic in the sheath.
This is just a consequence of how we determine the plate position from the upstream conditions (therefore, for dust transport, we still use Eq. (\ref{ie}),
without accounting for the electron drift velocity).
An important aspect in Figs. \ref{fig1} and \ref{fig1b} is that the plasma temperature decreases towards the plate.
This is due to the finite thermal conductivity of the plasma (particularly for the ions) and the fact that the heat flux must decrease while the plasma
is accelerated and the convective energy flux increases. Indeed Figs. \ref{fig1} and \ref{fig1b} (f) show that the total energy flux parallel to the plate is
conserved. The important observation is that the upstream energy flux is large and mostly parallel to the magnetic field ($\sim q_y$), but it decreases
substantially in the Chodura layer and the sheath. Thus, dust particles have a much better chance for survival in this environment if they can remain confined
near the walls.

\section{Dust transport simulations}\label{res}

In this section we study dust transport in the magnetized sheath-presheath discussed in Sect. \ref{sheathmodel}. 
We inject dust particles at the divertor plate with injection velocity $V_{xd}(t=0)=0.1$ m/s and follow their dynamics. The small injection velocity
is motivated by our focus on dust survivability and PMI: in order to withstand the harsh tokamak
environment, dust particles must necessarily move in proximity of the divertor plates/walls, where the energy fluxes are lower. 
A large injection velocity perpendicular to the plate,
as considered in other studies \cite{krash04, smirnov07}, will eject the dust particle outside the sheath-presheath, where 
for $q_0\simeq 10$ MW/m$^2$ it is likely
to sublimate or evaporate due to higher energy fluxes, and eventually redeposit non-locally.
The actual dust injection velocity in tokamaks is mostly an open question
(although it has been measured around $100$ m/s during disruptive events \cite{yu09}), reflecting the fact the dust generation mechanisms
(particularly for tungsten-based machines) are still poorly understood, and requires further experimental studies.
In general, for the parameters typical of long-pulse machines considered here, a micrometer tungsten particle with injection velocity
of a few $m/s$ remains confined in the sheath/Chodura layer, implying that our results are relevant to regimes that are not characterized
by fast and localized heat loads to the wall.

The physics of dust transport in the magnetized sheath is dominated by the electrostatic and ion drag forces, and 
can be described in the following terms \cite{tang10}:
as the dust particle is released into the plasma, it becomes charged and feels the effect of the electrostatic force.
For a negatively charged dust grain, the electrostatic force is directed away from the plate, since the plate is negatively charged,
and is stronger in the sheath.
At the same time the plasma flow exerts a drag force on the grain that tends to bring it back to the plate. The ion drag is stronger
away from the plate, where the plasma density is higher. Therefore,
in the direction perpendicular to the plate these two forces can balance and there can be an equilibrium position somewhere in the
sheath-presheath. Such equilibrium position depends on the dust radius, with smaller particles having equilibrium positions further away from the plate \cite{tang10}.
Thus, the dynamics perpendicular to the plate involves oscillatory motion around the equilibrium position and bouncing on the plate.
We note that for a positively charged grain there is no equilibrium position as both forces point towards the plate. In our simulations, however,
the dust particle can become positively charged only after an initial transient and the bouncing motion continues because of the dust inertia.
In the direction parallel to the plate, on the other hand, the ion drag force is unbalanced and the dust particle can travel long distances and be accelerated
to large velocities \cite{krash04}.

In order to characterize dust survivability and its impact on PMI, we consider a reference divertor
whose poloidal width is $L_{pol}=30$ cm, as done in Ref. \cite{tang10}. When a dust particle moves poloidally to a distance
greater than $L_{pol}$, it moves outside the divertor strike points to regions of the tokamak where the plasma is more benign.
Therefore we conclude that it survives, giving rise to a net loss of divertor material. When the poloidal transit distance is less than
$L_{pol}$, the dust particle does not survive and redeposits its material. 
In this case, when the characteristic poloidal distance traveled by the dust before destruction 
is less or comparable to the gyroradius of a tungsten atom sputtered by the wall
and promptly ionized ($\sim 300-500$ $\mu$m for $B_0=6$ T and a characteristic speed of a few eV), we conclude that redeposition is local.
Otherwise redeposition is non-local.
%We conclude that the dust particle redeposits
%non-locally when it travels relatively large distances (above mm) parallel to the plate, and consider
%smaller distances (below mm) as local redeposition. This classification is somewhat arbitrary, but it allows to draw some
%general conclusions on dust survivability and redeposition for particles of different size.

The dust parameters used in the simulations are mostly those used in DTOKS \cite{martin08, bacharis10}, apart for tungsten's specific heat capacity
where we choose $C_d=200$ J/kg K to account for the increase  at higher temperatures
relative to the value of $C_d=132$ J/kg K at room temperature and for tungsten emissivity, which is obtained by fitting 
the results of Ref. \cite{rosenberg08} for $r_d=0.1,\,1,\,10$ $\mu$m. We use a piece-wise linear fit of the dust temperature dependence between
$T_d=300$ K and $T_d=4000$ K, and assume that emissivity remains constant for $T_d>4000$ K. We do not model the dust size dependence of emissivity,
which affects the thermal radiation flux when the dust radius is reduced by evaporation.
Considering that above $T_d\sim 3000$ K emissivity increases  while the dust radius deacreases, this implies that our results are conservative in terms
of dust survival.
The dust parameters are reported in Table \ref{t1} for clarity. 
In addition, the temperature of the secondary emitted electrons is $T_{se}=3$ eV and the wall temperature is $T_{wall}=300$ K. We also
assume that collisions with the divertor plate are elastic.
We consider particles of different size centered around $r_d=1$ $\mu$m, which is the characteristic size of the dust
collected in current tokamaks \cite{jeff07}. In the event of significant sublimation or evaporation, we end the simulations when the dust radius
reaches $r_d=1$ nm, which is the limit where a continuous charging theory like OML begins to break down.

\begin{table}
\centering
\caption{Summary of the dust parameters used in the simulations of Sec. \ref{res}.}
\begin{tabular}{ccc}
\hline
\hline
       Property & Carbon & Tungsten\\
\hline
Density (kg/m$^3$) & $2250$ & $19300$\\
Specific heat capacity (J/kg/K) & $800$ & $200$\\
Melting temperature (K) & & $3695$\\
Sublimation temperature (K) & $3925$ & \\
Evaporation temperature (K) & & $5930$\\
Latent heat of fusion (J/kg) & & $1.92 \cdot 10^5$\\
Latent heat of sublimation (J/kg) &$2.97 \cdot 10^7$&\\
Latent heat of evaporation (J/kg) && $4.009 \cdot 10^6$\\
Thermionic work function (eV) & $5$ & $4.55$\\
$\delta_{max}$ & $1.0$ & $1.4$\\
$E_{max}$ (eV) & $300$ & $650$\\
Emissivity & $0.8$ & see text\\
\hline
\hline
\end{tabular}
\label{t1}
\end{table}

\subsection{Micrometer carbon dust particle transport in the magnetized sheath with $q_{0}\simeq1$ MW/m$^2$}

\begin{figure}
\centering
\includegraphics[scale=0.8]{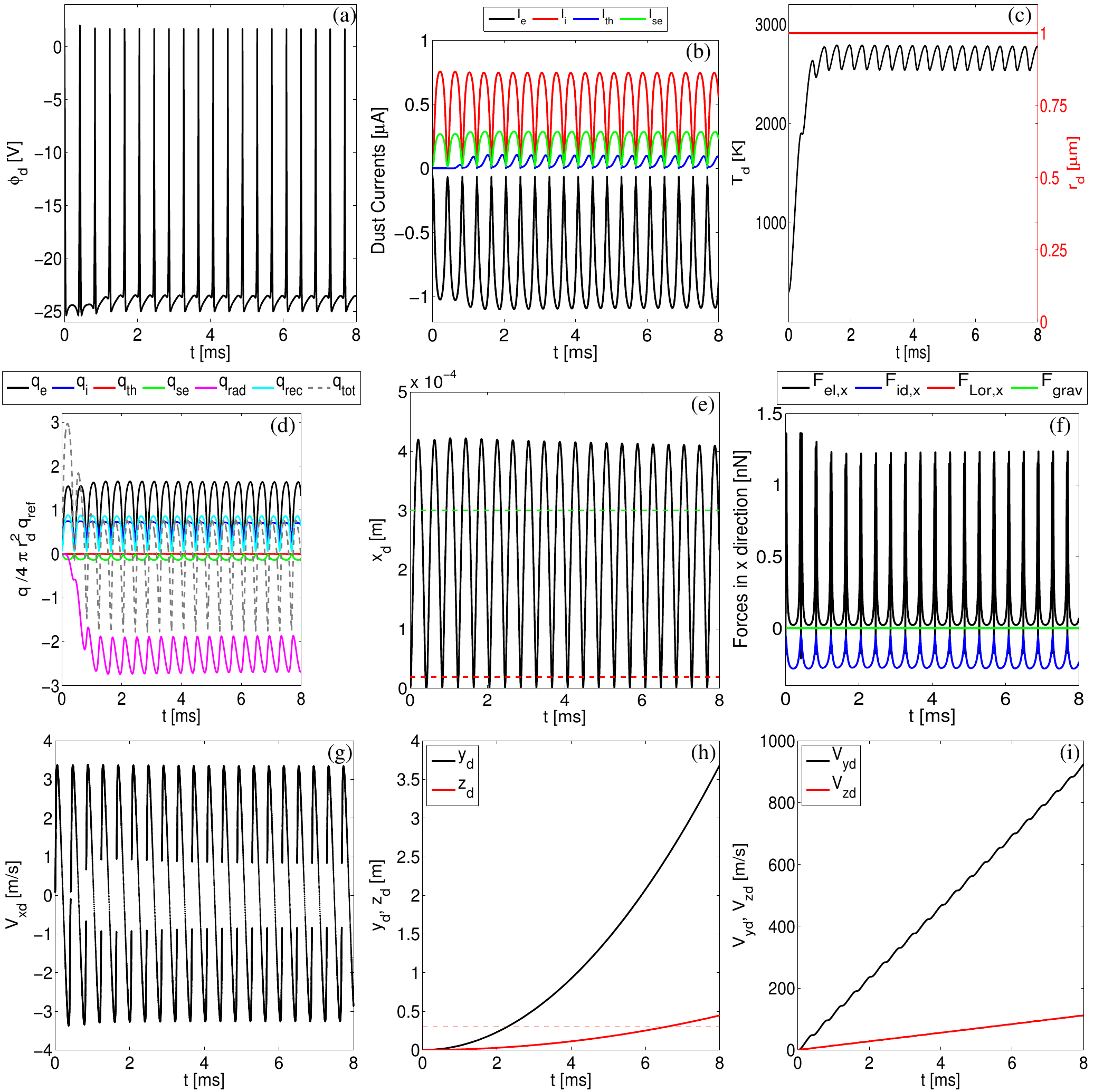}
\caption{(Color online) Dynamics of a carbon dust particle with $r_d=1$ $\mu$m in the sheath-presheath of a divertor plasma with $q_{0}=0.91$ MW/m$^2$.
Plotted quantities are: (a) dust potential, $\phi_d$; (b) dust currents; (c) dust temperature, $T_d$, and radius, $r_d$; (d) energy fluxes on the dust grain;
(e) dust position perpendicular to the plate, $x_d$; (f) forces acting on the dust grain in the direction perpendicular to the plate;
(g) dust velocity perpendicular to the plate, $V_{xd}$; (h) dust positions parallel to the plate, $y_d$ and $z_d$; dust velocities parallel to the plate,
$V_{yd}$ and $V_{zd}$. The dashed lines in Fig. (e)  correspond to the edges of the sheath (red) and of the Chodura layer (green), while
the dashed line in Fig. (h) corresponds to a reference divertor poloidal width $L_{pol}=30$ cm, and indicates whether the dust particle can transit across the whole
divertor poloidally.}
\label{fig2}
\end{figure}

Figure \ref{fig2} shows the evolution of model quantities for a simulation of dust transport for a $r_d=1$ $\mu$m
particle. The total energy flux in the direction perpendicular to the plate is $q_0=0.9$ MW/m$^2$ and we use carbon dust, therefore this case is 
representative of what a micrometer dust particle can experience in present-day short-pulse tokamaks, and allows
us to make contact with some results reported in the literature. The case of tungsten dust for the same reference parameters
is qualitatively similar and leads to the same conclusions (not shown).

As the dust particle is released from the plate, there is a very rapid
transient (on a time scale of tens of ns) during which the dust charges due
to the absorption of background plasma particles.  It is interesting
to note that in this particular instance the initial rapid charging
transient leads to a positively charged dust $\phi_d(t\simeq 5 \cdot
10^{-8} s)\simeq 1.8$ V. This is because very close to the plate the
electron density is much smaller than the ion density:
$n_i(x=0)/n_e(x=0)\simeq 29$. However, the initial injection velocity
is sufficient to move the dust towards regions of higher electron
density, where it quickly charges negatively and is then pushed away
from the plate by the electrostatic force.  Figure \ref{fig2} (a)
shows that the dust potential then oscillates between $-25.4$ V (away
from the plate) and $+2.0$ V (at the plate).  Figure \ref{fig2} (b)
shows the dust currents as a function of time. One can see that the
electron current is balanced by the ion and secondary electron
emission currents, while the thermionic current is less important
(particularly for $t<1$ ms). We note, however, that the oscillations
in the dust potential lose their initial symmetry precisely due to the
contribution of thermionic emission.  In general the plasma currents
are stronger away from the plate, when the plasma density is higher.
Figure \ref{fig2} (c) shows the time evolution of the dust temperature
and dust radius, while Fig. \ref{fig2} (d) shows the energy fluxes
from the various sources described in Sec. \ref{dustmodel}, normalized
to the reference value $q_{ref}=1$ MW/m$^2$.  The dust temperature
rises to about $T_d\simeq 2500$ K within the first ms, mainly due to
the energy fluxes associated with the background plasma and with the
recombination of electrons and ions on the dust grain surface.  As the
dust temperature rises, dust particle cooling due to thermal radiation
becomes important. Similarly to the dust currents in Fig. \ref{fig2}
(b), when the dust particle moves away from the plate the plasma has a
significantly higher density and delivers a net positive energy flux
to the dust.  On the other hand, the energy flux from the background
plasma becomes rather small in the sheath and the dust can cool by
thermal radiation.  Thus the dust temperature saturates, reaching a
maximum value of $T_d\sim 2800$ K well below the sublimation
temperature.  Consistently, the dust particle does not lose any mass
and its radius remains constant at $r_d=1$ $\mu$m, as shown in
Fig. \ref{fig2} (c).

Figure \ref{fig2} (e) shows the bouncing motion perpendicular to the plate that was discussed above. One can see that the dust particle
travels outside the Chodura layer, with a maximum excursion of about $4\rho_i$. The forces in the direction perpendicular to the plate
are shown in Fig. \ref{fig2} (f), indicating that indeed the electrostatic and ion drag forces are dominant in different parts of the sheath-presheath.
Figure \ref{fig2} (g) shows the dust velocity normal to the plate. It oscillates between $\pm 3.4$ m/s, indicating that the electrostatic force has accelerated
the particle well beyond the initial injection velocity of $0.1$ m/s.

Figure \ref{fig2} (h) shows the toroidal ($y_d$) and poloidal ($z_d$) distances traveled by the dust particle. At the end of the simulation,
the dust particles has moved $y_d\sim 3.7$ m toroidally and $z_d\sim45$ cm poloidally, which is larger than our reference poloidal divertor width
$L_{pol}=30$ cm. Thus, a dust particle
such as the one in Fig. \ref{fig2}, introduced for instance at the one edge of the divertor, could travel poloidally across the whole divertor and reach regions of the tokamak
where the plasma is more benign.
Figure \ref{fig2} (i) shows that the dust particle
can be accelerated to velocities of the order of several hundred m/s, particularly in the toroidal direction. This result is consistent with earlier simulation
results by Krasheninnikov and collaborators \cite{krash04,smirnov07}, and with experimental evidence of dust motion recorded from fast cameras 
in various tokamaks \cite{rudakov09}.
Interestingly, the fact that dust particles can travel long distances and experience a large number of collisions with the plate has been advocated
by Krasheninnikov \textit{et al.} as a mechanism to redirect dust motion towards the core, with important implications from the point
of view of core contamination \cite{krash04}. Key to this idea is the fact that the wall or divertor plates are characterized by some level
of surface roughness, whose characteristic length scale must be comparable with the dust grain radius to be effective.

In summary, for parameters relevant to current short-pulse tokamaks ($q_{0}\simeq1$ MW/m$^2$), we have confirmed that micrometer
dust particles born with small injection speeds remain confined to the sheath-presheath, are accelerated to high toroidal speeds, 
and can travel long distances in the tokamak chamber. These particles can survive near the walls rather easily, because they can cool
efficiently via thermal radiation. 

\subsection{Tungsten dust particle transport in the  magnetized sheath with $q_{0}\simeq10$ MW/m$^2$}

Next, we study dust transport in the divertor sheath-presheath with $q_0=9.6$ MW/m$^2$.
Since this case is relevant to long-pulse machines like ITER or DEMO, we only consider tungsten dust. 

\subsubsection{Micrometer dust particle}

\begin{figure}
\centering
\includegraphics[scale=0.8]{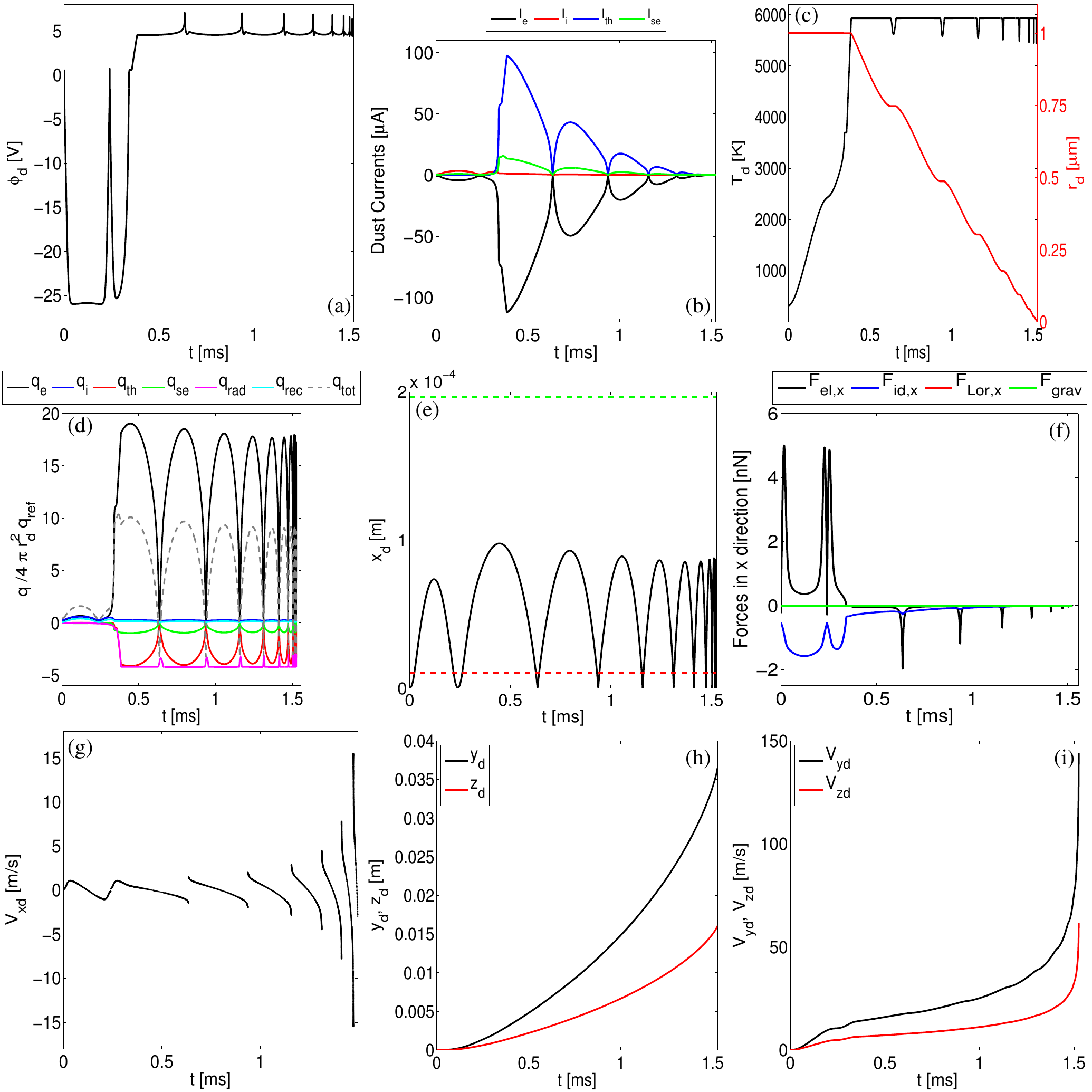}
\caption{(Color online) Dynamics of a tungsten dust particle with $r_d=1$ $\mu$m in the sheath-presheath of a divertor plasma with $q_{0}=9.6$ MW/m$^2$.
Plotted quantities are: (a) dust potential, $\phi_d$; (b) dust currents; (c) dust temperature, $T_d$, and radius, $r_d$; (d) energy fluxes on the dust grain;
(e) dust position perpendicular to the plate, $x_d$; (f) forces acting on the dust grain in the direction perpendicular to the plate;
(g) dust velocity perpendicular to the plate, $V_{xd}$; (h) dust positions parallel to the plate, $y_d$ and $z_d$; dust velocities parallel to the plate,
$V_{yd}$ and $V_{zd}$. The dashed lines in Fig. (e)  correspond to the edges of the sheath (red) and of the Chodura layer (green), while
the dashed line in Fig. (h) corresponds to a reference divertor poloidal width $L_{pol}=30$ cm, and indicates whether the dust particle can transit across the whole
divertor poloidally. Figure (g) shows only the first $98\%$ of the simulation and does not show the peak $V_{xd}\sim \pm569$ m/s.}
\label{fig3}
\end{figure}

Figure \ref{fig3} shows the evolution of model quantities
for a simulation of a micrometer dust particle.
Let us compare the dynamics of Fig. \ref{fig3} with that of Fig. \ref{fig2}.
Figure \ref{fig3} (a) shows the evolution of the dust potential. The initial dynamics is qualitatively similar to that of Fig. \ref{fig2} (a),
with the dust charging to a positive potential $\phi_d\simeq0.6$ V within $7$ ns, and then reaching $\phi_d\simeq -26$ V as it travels
away from the plate. However, during the second bounce on the plate, the evolution of the dust potential is very different from that in the first
bounce. This symmetry is broken by the fact that the dust temperature has already reached conditions that lead to a very strong thermionic
emission current, as it can be seen in Fig. \ref{fig3} (b). Unlike the case of Fig. \ref{fig2} (b), where the balance between the dust currents
is mainly due to the background electrons and ions collection for the entire simulation, in Fig. \ref{fig3} (b), after $t\approx 0.34$ ms,
this balance is mainly due to the background and thermionic emission electrons, with a relatively small contribution from the secondary
emitted electrons and practically none from the ion current. It is also worth emphasizing that thermionic emission induces a large spike
in the electron collection current, which peaks at about $I_e\approx -112$ $\mu$A.
Consistently, for $t>0.34$ ms the dust grain remains positively charged, with the dust
potential oscillating between $+4.4$ V and $+7$ V.
Figure \ref{fig3} (c) shows the evolution of the dust temperature and dust radius, while Fig. \ref{fig3} (d) shows the corresponding energy fluxes on the dust grain.
The dust temperature rises rather sharply: within $ t\approx 0.36$ ms the dust has melted completely ($T_d=3695$ K)
and at time $t \approx 0.39$ ms conditions for evaporation ($T_d=5930$ K) are met, the dust starts to lose mass and its radius shrinks. In $1.53$ ms
the dust radius reaches $r_d=1$ nm and the simulation ends. The energy fluxes in Fig. \ref{fig3} (d)
are consistent with the dust currents in Fig. \ref{fig3} (b). Initially the dust particle receives a strong positive net energy flux from the background plasma.
The rise of the dust temperature and of thermionic emission induces an energy flux collection spike from the background electrons, similar to what has
been observed in Ref. \cite{smirnov07}. Unlike the case of Fig. \ref{fig2} (d), the thermal radiation flux is unable to provide enough cooling, and the dust
particle evaporates. However, thermal radiation is still dominant when the dust particle is near the plate. This leads to the negative spikes in the dust
temperature and the corresponding plateau in the dust radius that can be seen in Fig. \ref{fig3} (c).

The dust motion in the direction perpendicular to the plate can be
seen in Fig. \ref{fig3} (e). Relative to Fig. \ref{fig2} (e), one can
see that the dust moves inside the sheath/Chodura layer, closer to the
plate. This is due to fact that in this case the background plasma has
$10$ times higher density and the ion drag is comparatively
stronger. While the dominant forces are still the ion drag and
electrostatic forces [Fig. \ref{fig3} (f)], for $t>0.34$ ms the dust
grain is positively charged, the electrostatic force points towards
the plate, and there is no equilibrium position in the
sheath-presheath. It is important to note, however, that
the number of bouncing collisions with the plate is reduced relative
to Fig. \ref{fig2}, and so is the probability for the dust particle to
be redirected by surface roughness.  The mass loss affects the
evolution of the dust velocity perpendicular to the plate, shown in
Fig. \ref{fig3} (g). During the first bounce the dust particle returns
to the plate with a velocity very similar to its injection velocity,
$V_{xd}(t\approx0.24\,{\rm ms})\approx -0.16$ m/s.  However, during
the second bounce, the electrostatic force starts pushing towards the
wall and the dust radius shrinks to $r_d(t\approx0.64\,{\rm
  ms})\approx 0.75$ $\mu$m, leading to $V_{xd}(t\approx0.64\,{\rm
  ms})\approx -1.5$ m/s. The process continues in a runaway fashion,
and the dust particle reaches a peak velocity perpendicular to the
plate of $\sim \pm 569$ m/s, about $160$ times larger than that
obtained in Fig. \ref{fig2} (e).  [We note that, for clarity of
Fig. \ref{fig3} (g), we have only plotted the first $98\%$ of the
simulation and the peak at $569$ m/s is not shown].

One more comment is in order with regard to the dynamics normal to the plate. For $t>0.39$ ms, the dust particle is a droplet and the issue of its
bouncing on the plate requires careful consideration. Indeed the impact of droplets on solid surfaces is a rather complex subject and a rich research
area (see for instance the review by Yaris \cite{yaris06}). Its outcome can be broadly classified in four categories: sticking, bouncing, spreading/deposition and
splashing, in order of increasing droplet velocity normal to the surface \cite{rein}. The thresholds between the various outcomes depend on many factors,
including physical and kinematic parameters of the droplet, as well as surface parameters like wettability and surface roughness \cite{yaris06,rein}.
It is customary to introduce the Weber number
${\rm We}=\rho_d V_{xd}^2 2 r_d/\gamma_d$,
where $\gamma_d$ is the dust surface tension (for tungsten, $\gamma_d=2.5$ N/m), which measures the strength of the droplet inertia normal to the surface to its surface tension.
Bouncing is typically associated with small ${\rm We}$ values \cite{rein}. For the simulation of Fig. \ref{fig3}, most of the dynamics occurs in this regime,
${\rm We}(t<1.51\,{\rm ms})\le 0.19$. In the last three bounces, however, the Weber number becomes of order unity because of the large dust
velocity attained by the dust particle normal to the plate and this could lead to spreading/deposition on the surface.
Furthermore, a liquid droplet in a plasma could in principle disrupt electrostatically.
This was studied in Ref. \cite{coppins10}, which shows that for tungsten dust particles in the low temperature plasmas considered here ($T_{e0}=10$ eV), electrostatic disruption
should not occur.

\begin{figure}
\centering
\includegraphics[scale=1]{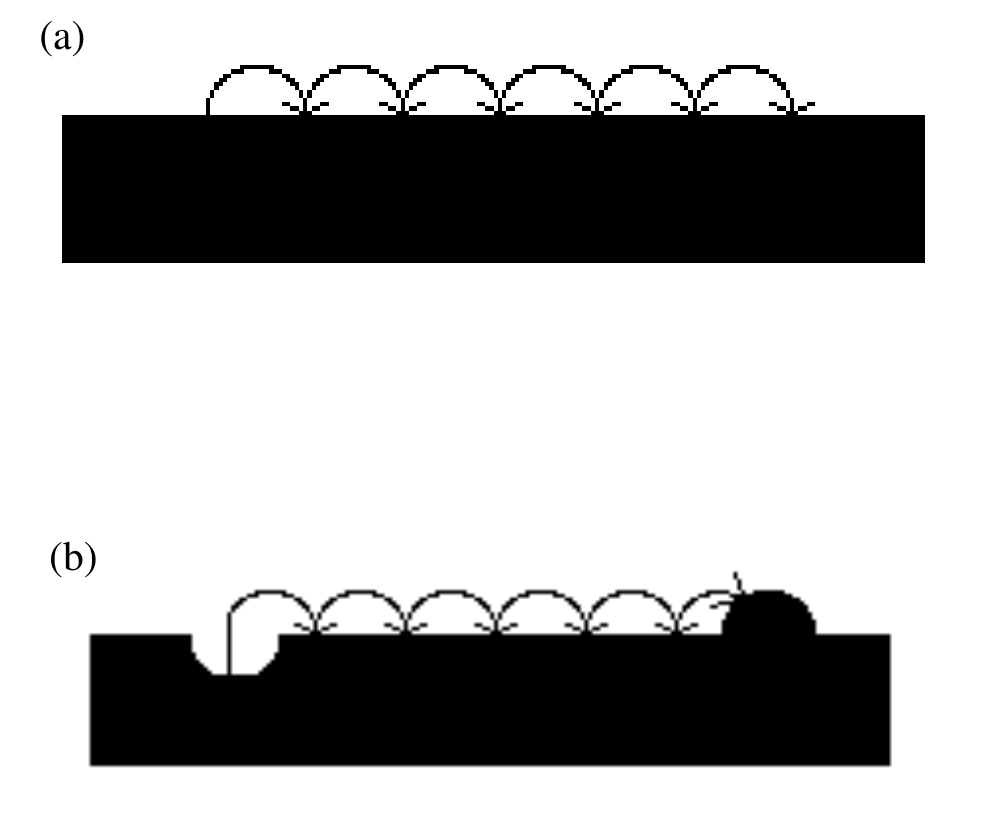}
\caption{Cartoon of the poloidal mass migration for micrometer dust particles: (a) the dust particle moves poloidally in the direction of the plasma flow,
(b) the non-local redeposition generates areas of net erosion and areas of net accumulation.}
\label{dune}
\end{figure}

The poloidal and toroidal transit distances traveled by the dust
particle are shown in Fig. \ref{fig3} (h) and the corresponding
velocities are in Fig. \ref{fig3} (i). The dust particle is still
accelerated to high speeds, although much more so towards the end of
the simulation, after the dust radius has shrunk considerably. The
particle has moved about $4$ cm toroidally, and about $2$ cm
poloidally at the end of the simulation.  The transit distances
obtained at the time in which the dust radius has shrunk to half its
size (and the dust particle has lost about $88\%$ of its mass),
$t\approx 0.9$ ms, are approximately $1$ cm toroidally and $6$ mm
poloidally.  Unlike the case of Fig. \ref{fig2}, the dust particle
evaporates before it can transit our reference divertor plate
poloidally.  The crucial observation, however, is that, owing to the
unbalanced ion drag acceleration in the poloidal direction, the dust
always move in the direction of the poloidal plasma flow and
redeposits its material non-locally. Over time, this leads to a net
\textit{poloidal mass migration} from one side of the divertor plate
to the other, which bears some resemblance with the formation and
transport of sand dunes in the desert, shaped by the wind flow. A
cartoon of the poloidal mass migration is sketched in Fig. \ref{dune}.
In order to transit the reference divertor poloidally without
evaporating, the dust particle should have a poloidal injection speed
$V_{zd}(t=0)\ge 770$ m/s. This sets a design constraint for the dust
shield concept in Ref.~\cite{tang10}.

In summary, we have shown that the dynamics of a micrometer dust
particle near the divertor plates of a tokamak characterized by an
energy flux perpendicular to the plate $q_{0}\simeq10$ MW/m$^2$ is
completely different from that in a plasma with $q_{0}\simeq1$
MW/m$^2$: the survivability and transport of micrometer dust particles
are drastically reduced. The dust particle redeposits its material
non-locally, with a net poloidal mass migration from one side of the
divertor plate to the other.

\subsubsection{Sub-micrometer dust particle}

Next, we analyze the dynamics of a tungsten dust particle with radius $r_d=0.1$ $\mu$m (all other parameters are unchanged). 

\begin{figure}
\centering
\includegraphics[scale=0.8]{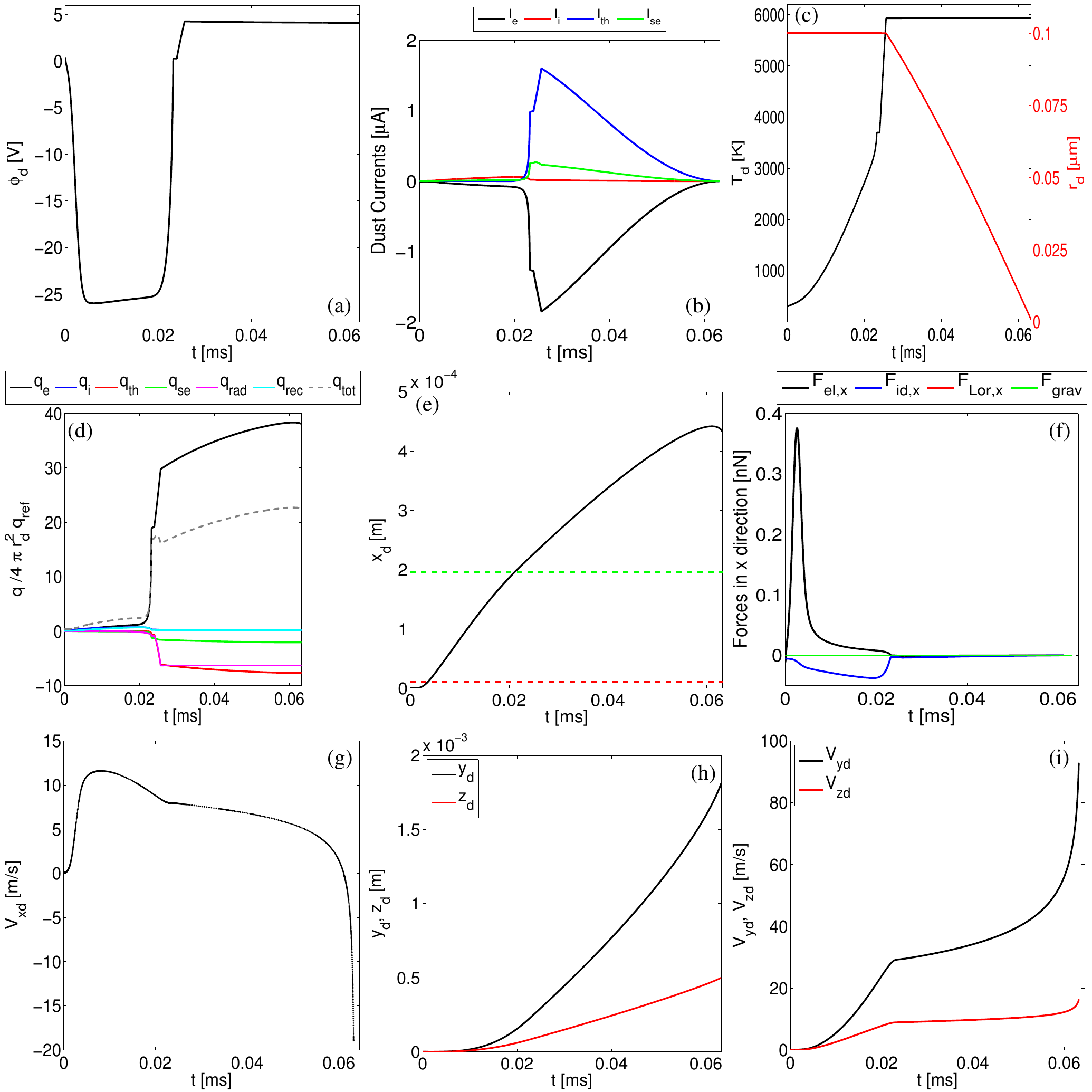}
\caption{(Color online) Dynamics of a tungsten dust particle with $r_d=0.1$ $\mu$m in the sheath-presheath of a divertor plasma with $q_{0}=9.6$ MW/m$^2$.
Plotted quantities are: (a) dust potential, $\phi_d$; (b) dust currents; (c) dust temperature, $T_d$, and radius, $r_d$; (d) energy fluxes on the dust grain;
(e) dust position perpendicular to the plate, $x_d$; (f) forces acting on the dust grain in the direction perpendicular to the plate;
(g) dust velocity perpendicular to the plate, $V_{xd}$; (h) dust positions parallel to the plate, $y_d$ and $z_d$; dust velocities parallel to the plate,
$V_{yd}$ and $V_{zd}$. The dashed lines in Fig. (e)  correspond to the edges of the sheath (red) and of the Chodura layer (green), while
the dashed line in Fig. (h) corresponds to a reference divertor poloidal width $L_{pol}=30$ cm, and indicates whether the dust particle can transit across the whole
divertor poloidally.}
\label{fig4}
\end{figure}

The evolution of the model quantities is shown
in Fig. \ref{fig4} with the same format of Figs. \ref{fig2} and \ref{fig3}. 
Figure \ref{fig4} (a) shows the evolution of the dust potential. The initial part of the dynamics is qualitatively similar to that
of Figs. \ref{fig2} and \ref{fig3}, with the dust reaching a negative potential of about $\phi_d\approx-26$ V in $t\approx 6$ $\mu$s.
At this point the dust starts to heat up and the thermionic current becomes dominant together with
the background electron current [Fig. \ref{fig4} (b)]. At $t\approx 23$ $\mu$s the dust becomes positively
charged, and for $t>26$ $\mu$s the dust potential saturates to $\phi_d\approx +4.2$V. Figure \ref{fig4} (c) shows the evolution of the 
dust temperature and radius. Melting conditions are reached at $t\approx 23$ $\mu$s and the dust starts to evaporate at $t\approx 26$ $\mu$s.
Within $t\approx 63$ $\mu$s the dust particle has shrunk to $r_d=1$ nm, much faster than the case of Fig. \ref{fig3}, where it took $t\approx 1.53$ ms
for a micrometer dust particle to reach nanometer size, in part owing to the initial size difference. 
The energy fluxes are shown in Fig. \ref{fig4} (d): the dust particle is unable to cool by
thermal radiation or by the thermionic energy flux and as a result the total energy flux is always positive and large,
dominated by the thermionically-induced energy flux collection spike from the background electrons.
 Consistently, when evaporation
conditions are reached, the dust completely evaporates away and the dust temperature remains constant for the rest of the simulation, unlike the 
case of Fig. \ref{fig3} (c) where the dust particle can still cool when it travels in proximity of the plate.

Figure \ref{fig4} (e) shows the time evolution of the dust position perpendicular to the plate. 
As we discussed above, in the direction perpendicular to the plate 
smaller dust particles are characterized by an equilibrium position (where the electrostatic force and the ion drag balance) further away from the plate
and this is shown clearly in Fig. \ref{fig4} (e): the dust particle is accelerated outside the sheath/Chodura layer by the electrostatic force. 
From Fig. \ref{fig4} (f), showing the time evolution of various forces acting on the dust particle, one can see that the ion drag force begins to
dominate at $t\approx 8$ $\mu$s but, because of inertia, only at $t\approx 61$
$\mu$s the dust particle inverts its motion to return towards the plate. 
From the perspective of the dust particle survivability, however,
this is too late. Traveling outside the sheath/Chodura layer means facing higher energy fluxes and the dust particle evaporates before returning to the plate.
From Fig. \ref{fig4} (g), one can see that the electrostatic force accelerates the particle to about $12$ m/s.

At the end of the simulation, the dust particle has traveled approximately $2$ mm toroidally and $0.5$ mm poloidally [Fig. \ref{fig4} (h)],
while it gets accelerated to about $93$ m/s toroidally and $16$ m/s poloidally [Fig. \ref{fig4} (i)].
The transit distances corresponding to the time where the dust has lost $\sim 88 \%$ of its mass, $t \approx 46$ $\mu$s, are about $1$ mm toroidally
and $0.3$ mm poloidally, respectively. Clearly the dust particle does not move much, and cannot transit across our reference divertor poloidally to find safer regions
of the tokamak. Although as in the case of a micrometer dust particle the mass redeposition occurs in the direction of the plasma flow, 
most of the dust mass is lost within a distance comparable to that of a promptly redeposited sputtered tungsten ion and we classify this case as local redeposition.
The critical poloidal injection velocity to transit the reference divertor poloidally without reaching evaporation
conditions is $V_{zd}(t=0)\simeq 12$ km/s.

In summary, a sub-micrometer dust particle cannot survive in a tokamak divertor plasma characterized by an energy flux perpendicular to the plate
$q_{0}\simeq10$ MW/m$^2$ and redeposits its material locally.

\subsubsection{Supra-micrometer dust particle}

Next, we study the dynamics of a tungsten dust particle with $r_d=10$ $\mu$m (all the other parameters are unchanged).
We note that our model treats the dust particle as a point particle,
therefore its application for bigger particles that tend to move closer to the plate is somewhat marginal. Nevertheless, we will still use it
to get some sense on dust transport in this parameter regime.

\begin{figure}
\centering
\includegraphics[scale=0.8]{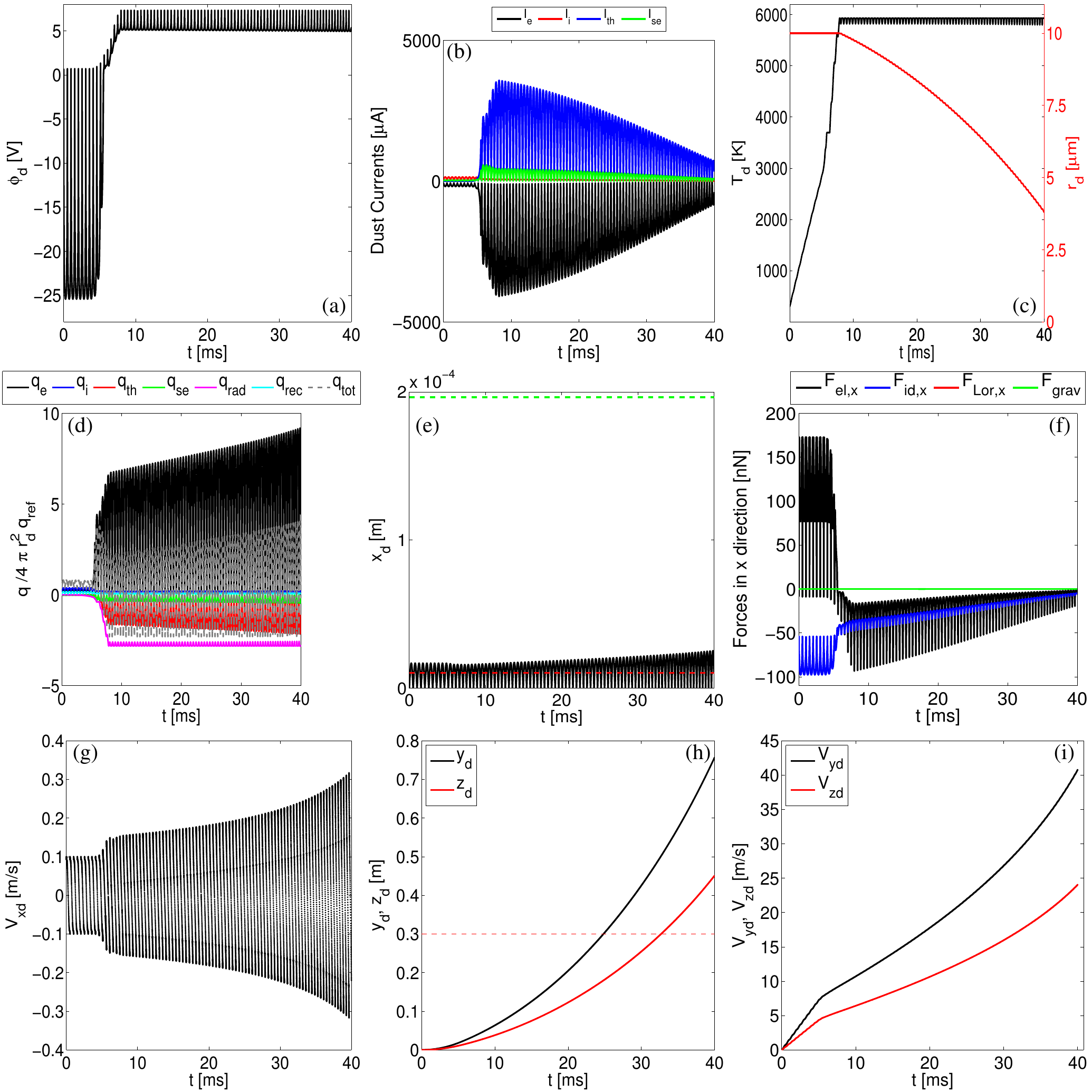}
\caption{(Color online) Dynamics of a tungsten dust particle with $r_d=10$ $\mu$m in the sheath-presheath of a divertor plasma with $q_{0}=9.6$ MW/m$^2$.
Plotted quantities are: (a) dust potential, $\phi_d$; (b) dust currents; (c) dust temperature, $T_d$, and radius, $r_d$; (d) energy fluxes on the dust grain;
(e) dust position perpendicular to the plate, $x_d$; (f) forces acting on the dust grain in the direction perpendicular to the plate;
(g) dust velocity perpendicular to the plate, $V_{xd}$; (h) dust positions parallel to the plate, $y_d$ and $z_d$; dust velocities parallel to the plate,
$V_{yd}$ and $V_{zd}$. The dashed lines in Fig. (e)  correspond to the edges of the sheath (red) and of the Chodura layer (green), while
the dashed line in Fig. (h) corresponds to a reference divertor poloidal width $L_{pol}=30$ cm, and indicates whether the dust particle can transit across the whole
divertor poloidally.}
\label{fig5}
\end{figure}

The evolution of the dust potential is shown in Fig. \ref{fig5} (a), complemented by the dust currents in Fig. \ref{fig5} (b).
Qualitatively the dynamics resembles that of Figs. \ref{fig3} and \ref{fig4}. In the initial part of the dynamics the dust particle
is mostly negatively charged, with a negative peak of $\phi_d\approx -25.4$ V and the current balance occurs mainly through
the collection of background electrons and ions. As the dust particle heats up, the later part of the dynamics is strongly
influenced by thermionic emission: for $t>5.5$ ms the dust remains positively charged and the current balance is mainly through
the background electron collection and thermionic emission. The evolution of the dust temperature and of the dust radius is shown
in Fig. \ref{fig5} (c). Initially the temperature rises sharply, with melting conditions reached at $t\approx 5.8$ ms and complete melting
achieved at $t\approx 6.3$ ms. After $t\sim 8$ ms, the dust particle starts to evaporate and the temperature exhibits oscillations associated
with the bouncing motion, similar to those in Fig. \ref{fig3}. The dust radius shrinks to about $4$ $\mu$m by the end of the simulation.
The energy fluxes on the dust
grain are shown in Fig. \ref{fig5} (d), showing the electron heat flux collection spike induced by thermionic emission as the main mechanism
heating the dust particle, contrasted by thermal radiation and the thermionic emission energy flux as the main cooling mechanisms.
As in Fig. \ref{fig3} (d), the dust particle is able to cool when moving close to the plate. In general, however, the energy fluxes normalized to
$4 \pi r_d^2 q_{ref}$ are comparatively smaller for the bigger dust particle.

Figure \ref{fig5} (e) shows the dust motion perpendicular to the plate. The motion is mostly concentrated in the sheath since, as we have argued above,
bigger particles have equilibrium position between the electrostatic force and the ion drag closer to the plate. Bigger excursions
inside the Chodura layer occur in the later part of the dynamics, when thermionic emission becomes important. The forces acting on the dust
particle are shown in Fig. \ref{fig5} (f). The velocity perpendicular to the plate, shown in Fig. \ref{fig5} (g), rises slowly to a peak of approximately $\pm0.3$ m/s.
The Weber number remains small for the whole simulation, ${\rm We}<0.01$, indicating that the dust droplet can bounce on the divertor plate. 

Figure \ref{fig5} (h) shows the dust transit distances parallel to the plate: at the end of the simulation the dust particle has traveled about $76$ cm toroidally and
$45$ cm poloidally. This indicates that the dust particle can transit across our reference poloidal divertor length to move out of the divertor strike points, and
survive. By the time this happens, $t\sim 33$ ms, the dust radius is $r_d\simeq 5.6$ $\mu$m corresponding to a $\sim 80 \%$ mass loss redeposited non-locally.
Because of the bigger size, the dust velocity parallel to the plate is much smaller than that of the micrometer particle studied in Fig. \ref{fig3},
peaking at $33$ m/s toroidally and $19$ m/s poloidally [Fig. \ref{fig5} (i)]. 

In summary, supra-micrometer dust particles with $r_d\sim10$ $\mu$m in a divertor plasma characterized by an energy flux perpendicular to the plate
$q_{0}\simeq10$ MW/m$^2$ can survive partially, therefore causing a net loss of divertor material and some non-local redeposition.

\section{Conclusions} \label{concl}

We have studied dust transport in a tokamak plasma. Our model includes a dust charging equation, the dust equation of motion, the dust heating equation
and an equation for dust mass loss in the framework of the OML theory, 
and is similar to other dust transport models used in the literature \cite{ticos06,pigarov05,smirnov07,martin08,bacharis10,ratynskaia13}.

Our focus is on dust survivability and its impact on the PMI problem. Therefore we have considered small dust injection velocities perpendicular to the divertor
plate, and the dust particles remain confined in the magnetized sheath where the energy fluxes are lower. The dust transport model is therefore coupled
with a (one dimensional) model of the magnetized sheath which stems from the Braginskii fluid equations. It features an equation for the conservation of the
total energy flux perpendicular to the plate, which is used to derive sheath profiles consistent with a prescribed energy flux upstream.
We have studied two classes of magnetized sheath equilibria. The first is characterized by the upstream energy flux $q_{0}\simeq 1$ MW/m$^2$, and is relevant
to current short-pulse tokamaks. The second has $q_{0}\simeq 10$ MW/m$^2$, and is relevant to next generation long-pulse tokamaks like ITER or DEMO.

We have shown that micrometer dust particles can survive with relative ease near the divertor plates when $q_{0}\simeq 1$ MW/m$^2$, since they can cool
efficiently by thermal radiation. The situation is completely different when $q_{0}\simeq 10$ MW/m$^2$, where the survivability and transport of micrometer
particles are drastically reduced. Specifically, for $q_{0}\simeq 10$ MW/m$^2$ and small dust injection velocities, we have shown that
\begin{itemize}

\item small size dust particles ($r_d\sim 0.1$ $\mu$m) cannot survive and redeposit their material locally;

\item medium size dust particles ($r_d\sim 1$ $\mu$m) mostly cannot survive and redeposit their material non-locally. Since the dust always moves in the
direction of the poloidal plasma flow, this leads to a poloidal net mass migration across the divertor;

\item large size dust particles ($r_d\sim 10$ $\mu$m) can survive partially, leading to a net loss of divertor material and non-local redeposition.

\end{itemize}
The dust injection speed is obviously critical for survival. A large injection velocity normal to the plate leads to destruction, as the dust particle meets higher energy fluxes.
A large poloidal injection velocity improves the chances for survival. Our calculations indicate that poloidal velocities of the order of $1$ km/s are necessary for
a tungsten microparticle to transit across the divertor poloidally without evaporating.

The picture just described needs to be complemented by the material science perspective to provide the characteristic dust size and injection speed relevant
to long-pulse tokamaks, but can be used to draw some general conclusions.
For instance, if the characteristic dust size generated in ITER is $r_d\sim 1$ $\mu$m, our study indicates that large quantities of dust
should not be accumulated in the machine and the dust safety limits might not be a problem. On the other hand, the non-local redeposition and the related
net poloidal mass migration suggest that matching local erosion and redeposition profiles might be a challenge. A characteristic dust size of $r_d\sim 10$ $\mu$m
is possibly worse, since large quantities of dust could potentially be accumulated in addition to the challenge of matching local erosion and redeposition profiles.

\medskip{}
\acknowledgements 
This work was funded by the U.S. Department of Energy Office of
Science, Office of Fusion Energy Sciences, under the auspices of the
National Nuclear Security Administration of the U.S. Department of
Energy by Los Alamos National Laboratory, operated by Los Alamos
National Security LLC under contract DE-AC52-06NA25396.

\bibliography{braginskii_revised2_arxiv}

\pagebreak

\pagebreak

\pagebreak

\pagebreak

\pagebreak

\pagebreak

\pagebreak

\end{document}